\documentclass[12pt]{article}

\parskip=\medskipamount

\usepackage{tensor}
\usepackage{ulem}
\usepackage{physics}
\usepackage{amsmath,amssymb,calc, amsthm,bbm, epsfig,psfrag, mathtools}
\newtheorem{theorem}{Theorem}

\usepackage{latexsym,bm,amsfonts}
\usepackage{graphicx, enumerate}
\usepackage{enumitem}
 \usepackage[all,cmtip]{xy}
\usepackage[numbers,sort&compress]{natbib}
\usepackage[dvipsnames]{xcolor}
\usepackage{mathrsfs}
\usepackage{tikz,tikz-cd}

\numberwithin{equation}{section}

\allowdisplaybreaks

\newcommand{\ul}{\underline}

\newcommand{\Comment}[1]{{}}
\definecolor{darkblue}{rgb}{0.15,0.35,0.55}
\definecolor{reddish}{rgb}{0.65, 0.2, 0.2}
\usepackage[linktocpage=true]{hyperref}
\hypersetup{
colorlinks=true,
citecolor=darkblue,
linkcolor=reddish,
urlcolor=darkblue,
pdfauthor={},
pdftitle={},
pdfsubject={}
}

\makeatletter
\renewcommand\section{\@startsection {section}{1}{\z@}%
                                   {-3.5ex \@plus -1ex \@minus -.2ex}
                                   {2.3ex \@plus.2ex}%
                                   {\normalfont\large\bfseries}}
\renewcommand\subsection{\@startsection{subsection}{2}{\z@}%
                                     {-3.25ex\@plus -1ex \@minus -.2ex}%
                                     {1.5ex \@plus .2ex}%
                                     {\normalfont\bfseries}}
\makeatother

\newfont{\goth}{ygoth.tfm scaled 1200}                   

\newcommand{\overbar}[1]{\mkern 1.5mu\overline{\mkern-1.5mu#1\mkern-1.5mu}\mkern 1.5mu}

\def\TT{{T\overbar{T}}}

\def\cO{{\mathcal{O}}}
\def\fO{{\mathfrak{O}}}

\begin{document}
\begin{titlepage}
\begin{flushright}
\today
\end{flushright}
\vspace{5mm}

\begin{center}
{\Large \bf 
Quantization of the ModMax Oscillator
}
\end{center}

\begin{center}

{\bf
Christian Ferko,${}^{a}$
Alisha Gupta,${}^{b}$
and
Eashan Iyer${}^{b}$
} \\
\vspace{5mm}

\footnotesize{
${}^{a}$
{\it 
Center for Quantum Mathematics and Physics (QMAP), 
\\ Department of Physics \& Astronomy,  University of California, Davis, CA 95616, USA
}
 \\~\\
${}^{b}$
{\it 
The Academy for Mathematics, Science, and Engineering \\ Morris Hills High School, Rockaway, NJ 07866, USA
}}
\vspace{2mm}
~\\
\texttt{caferko@ucdavis.edu, alishag0101@gmail.com, eashanriyer@gmail.com}\\
\vspace{2mm}

\end{center}

\begin{abstract}
\baselineskip=14pt

\noindent We quantize the ModMax oscillator, which is the dimensional reduction of the Modified Maxwell theory to one spacetime dimension. We show that the propagator of the ModMax oscillator satisfies a differential equation related to the Laplace equation in cylindrical coordinates, and we obtain expressions for the classical and quantum partition functions of the theory. To do this, we develop general results for deformations of quantum mechanical theories by functions of conserved charges. We show that canonical quantization and path integral quantization of such deformed theories are equivalent only if one uses the phase space path integral; this gives a precise quantum analogue of the statement that classical deformations of the Lagrangian are equivalent to those of the Hamiltonian.


\end{abstract}
\vspace{5mm}

\vfill
\end{titlepage}

\newpage
\renewcommand{\thefootnote}{\arabic{footnote}}
\setcounter{footnote}{0}

\tableofcontents{}
\vspace{1cm}
\bigskip\hrule


\allowdisplaybreaks

\section{Introduction}\label{sec:intro}

Historically, quantum field theories first arose via the quantization of classical field theories. Following the modern usage of the term \cite{Poland:2022zhe}, we understand a quantum field theory (QFT) to mean any model that is compatible with certain physical principles including quantum mechanics, locality, and Lorentz invariance on a fixed $(d+1)$-dimensional spacetime manifold. When $d = 0$, the Lorentz structure becomes essentially trivial and one has an ordinary theory of quantum mechanics.

However, we still lack a systematic understanding of the process of quantization for several reasons. One reason is that it is not known how to uniquely quantize a general classical theory, except in the case of theories which can be brought into a conventional form with a quadratic kinetic term. A famous example is the Nambu-Goto action of string theory; rather than attempting to quantize this theory directly, one first rewrites it in the classically equivalent form of the Polyakov action, which can then be quantized because the theory is quadratic in derivatives. A second reason is that not all quantum field theories admit classical limits. This means that certain QFTs cannot ever be understood by quantization of a classical theory, defined for instance by a Lagrangian or Hamiltonian (indeed, many QFTs are non-Lagrangian and thus do not even admit such a description). Because of these observations, it is sometimes said that ``quantization is not a functor.''

In order to better understand quantization and the space of QFTs, it seems that one must develop new tools. One such tool is to describe new quantum field theories using controlled deformations of old ones. An example of such a deformation, which has generated considerable interest in the past several years, is the $\TT$ deformation of two-dimensional QFTs. The $\TT$ operator refers to the coincident point limit
\begin{align}\label{TT_defn}
    \mathcal{O}_{\TT} ( x ) = \lim_{y \to x} \left( T^{\mu \nu} ( x ) T_{\mu \nu} ( y ) - \tensor{T}{^\mu_\mu} ( x ) \tensor{T}{^\nu_\nu} ( y ) \right) \, , 
\end{align}
which was shown in \cite{Zamolodchikov:2004ce} to define a local operator in any translation-invariant $2d$ QFT.

Using any such $2d$ QFT as a seed theory, one can define a family of theories, labeled by a flow parameter $\lambda$, which arise from deforming the seed theory by $\TT$. At the classical level, we think of this parameterized family of actions as solving the flow equation
\begin{align}\label{2d_TT_defn}
    \frac{\partial S_\lambda}{\partial \lambda} = \frac{1}{2} \int d^2 x \, \left( T^{(\lambda) \mu \nu} T^{(\lambda)}_{\mu \nu} - \left( \tensor{T}{^{(\lambda)}^\mu_\mu} \right)^2 \right) \, ,
\end{align}
where $T_{\mu \nu}^{(\lambda)}$ is the stress tensor computed from $S_\lambda$,
\begin{align}\label{hilbert_stress}
    T_{\mu \nu} = \frac{-2}{\sqrt{-g}} \frac{\delta S_\lambda}{\delta g^{\mu \nu}} \, .
\end{align}
However, the interpretation of the differential equation (\ref{2d_TT_defn}) for the classical Lagrangian can be somewhat subtle. Because $\mathcal{O}_{\TT}$ exists in the spectrum of local operators in the seed theory, deforming by this operator should lead to a well-defined quantum theory. One could ask whether this quantum theory corresponds to the quantization, in some appropriate sense, of the classical action $S_{\lambda}$.

To address this question, one can use an alternative characterization of the quantum theory obtained by a $\TT$ deformation. For instance, it is known \cite{Dubovsky:2012wk,Dubovsky:2013ira} that the S-matrix for scattering in a $\TT$-deformed QFT is obtained by dressing the S-matrix of the undeformed theory with a momentum-dependent phase known as a CDD factor \cite{PhysRev.101.453}. This gives an independent description of scattering in the quantum theory which can be compared to predictions from quantization of the solution to (\ref{2d_TT_defn}). At one-loop level, one must add specific counter-terms when renormalizing the classical Lagrangian in order to reproduce the expected behavior of the $\TT$-deformed S-matrix \cite{Rosenhaus:2019utc,Dey:2021jyl,Chakrabarti:2022lnn}.\footnote{Because the $\TT$ operator is irrelevant in the sense of the renormalization group, this behavior is partly expected. Adding a generic irrelevant operator will typically activate infinitely many counterterms; the surprise is that the irrelevant $\TT$ deformation does not lead to a loss of analytic control.} This suggests that the action which solves equation (\ref{2d_TT_defn}) does not, by itself, contain enough data to quantize and obtain the correct $\TT$-deformed QFT at the quantum level; additional information from the S-matrix characterization is needed.

Another piece of evidence for this perspective comes from the observation that one can use different notions of the energy-momentum tensor in defining the flow (\ref{2d_TT_defn}). For instance, the Noether stress tensor is defined as the conserved current associated with spatial translations, while the Hilbert stress tensor (\ref{hilbert_stress}) is defined as the variation of the action with respect to the metric; these two notions do not agree in general, and one can also consider other stress tensors which are related to these two by improvement transformations. For theories involving fermions, a direct quantization of the classical actions which solve the flow equations (\ref{2d_TT_defn}) driven by different definitions of the stress tensor leads to inequivalent Hilbert spaces \cite{Lee:2021iut,Lee:2023uxj}.

We will interpret these observations by taking the following perspective. Although the solution to the classical $\TT$ flow equation (\ref{2d_TT_defn}) is useful, and often gives interesting hints about the nature of a $\TT$-deformed QFT, the process of quantizing this deformed Lagrangian can be ambiguous. Indeed, as we have stressed above, quantization is not a functor: except in simple cases such as free theories, we do not understand a unique and systematic prescription for turning classical theories into quantum ones. Rather, what we \emph{mean} by the quantum theory of a $\TT$-deformed seed is determined by other characterizations such as the S-matrix or torus partition function \cite{Cardy:2018sdv,Datta:2018thy,Aharony:2018bad}. These independent pieces of data should be viewed as picking out the correct prescription for performing the quantization of the $\TT$-deformed Lagrangian. This is in analogy with the viewpoint that the proper quantization prescription for the Nambu-Goto string is the one which proceeds by first rewriting the theory in Polyakov form and then quantizing using the path integral.

It is natural to ask whether adopting this perspective offers us insights into the quantization of other models. Recently, a family of related theories which exhibit non-analytic square-root structures in their Lagrangians have been introduced, all of which satisfy some classical flow equation similar to (\ref{2d_TT_defn}). We will now take a detour to describe some purely classical aspects of this collection of theories before returning to issues of quantization.

The first member of this class to be introduced was a four-dimensional gauge theory known as the Modified Maxwell or ModMax model \cite{Bandos:2020jsw}, which is described by the action
\begin{align}\label{modmax_lagrangian}
    S_{\text{ModMax}} ( \gamma ) = \frac{1}{4} \int d^4 x \, \left( 
    - \cosh ( \gamma ) F^{\mu \nu} F_{\mu \nu}
    + \sinh ( \gamma ) 
    \sqrt{ 
    \left( F^{\mu \nu} F_{\mu \nu} \right)^2 + \left( F^{\mu \nu} \widetilde{F}_{\mu \nu} \right)^2
    } \right) \, , 
\end{align}
where $F_{\mu \nu}$ is the field strength of the Abelian gauge field $A_\mu$ and $\widetilde{F}^{\mu\nu}= \frac{1}{2}\varepsilon^{\mu\nu\rho\tau}F_{\rho\tau}$ is its Hodge dual. When $\gamma = 0$, the action (\ref{modmax_lagrangian}) reduces to that of the usual Maxwell theory.

As a classical theory, the ModMax model (\ref{modmax_lagrangian}) exhibits several intriguing properties. It is the unique conformally invariant and electric-magnetic duality-invariant extension of the Maxwell theory. It also satisfies a flow equation driven by a function of the energy-momentum tensor \cite{Babaei-Aghbolagh:2022uij,Conti:2022egv}, namely
\begin{align}\label{root_TTbar_flow_4d}
    \frac{\partial S_{\text{ModMax}} ( \gamma ) }{\partial \gamma} = \frac{1}{2} \int d^4 x \, \sqrt{ T^{(\gamma) \mu \nu} T_{\mu \nu}^{(\gamma)} } \, ,
\end{align}
where $T_{\mu \nu}^{(\gamma)}$ is the stress tensor of the ModMax theory (\ref{modmax_lagrangian}) at parameter $\gamma$. Unlike the flow (\ref{2d_TT_defn}) for the Lagrangian of a $\TT$-deformed $2d$ QFT, the operator on the right side of (\ref{root_TTbar_flow_4d}) is classically marginal. Note that, since the ModMax theory is conformally invariant and thus its stress tensor has vanishing trace, the operator driving the flow (\ref{root_TTbar_flow_4d}) need not have any dependence on $\tensor{T}{^{(\gamma)}^\mu_\mu}$. However, it is convenient to define another combination which does involve the trace and which reduces to (\ref{root_TTbar_flow_4d}) in the conformal limit. For a theory in $D$ spacetime dimensions with energy-momentum tensor $T_{\mu \nu}$, let
\begin{align}
    \mathcal{R}^{(D)} = \sqrt{ \frac{1}{D} T^{\mu \nu} T_{\mu \nu} - \frac{1}{D^2} \left( \tensor{T}{^\mu_\mu} \right)^2 } \, .
\end{align}
In terms of the traceless part of the stress tensor, which we write as $\widehat{T}_{\mu \nu} = T_{\mu \nu} - \frac{1}{D} g_{\mu \nu} \tensor{T}{^\rho_\rho}$, this operator is simply
\begin{align}
    \mathcal{R}^{(D)} = \frac{1}{\sqrt{D}} \sqrt{ \widehat{T}^{\mu \nu} \widehat{T}_{\mu \nu} } \, .
\end{align}
Including this dependence on the trace allows us to extend certain flow equations to non-conformal theories. For instance, there is a two-parameter family of ModMax-Born-Infeld theories labeled by couplings $(\lambda, \gamma)$, which reduces to (\ref{modmax_lagrangian}) when $\lambda = 0$ and to the Born-Infeld theory when $\gamma = 0$. This family satisfies two commuting classical flow equations, one driven by a four-dimensional version of the $\TT$ operator and one driven by the operator $\mathcal{R}^{(4)}$ \cite{Conti:2018jho,Ferko:2022iru,Ferko:2023ruw}. The operator $\mathcal{R}^{(3)}$ also appears in the flow equation which deforms the $3d$ Maxwell Lagrangian into the Born-Infeld theory in three dimensions \cite{Ferko:2023sps}.

When $D = 2$, the combination $\mathcal{R}^{(2)}$ is the root-$\TT$ operator introduced in \cite{Ferko:2022cix}.\footnote{We refer the reader to \cite{Rodriguez:2021tcz,Bagchi:2022nvj,Tempo:2022ndz,Hou:2022csf} for other work related to the root-$\TT$ operator.} Applying this deformation to the seed theory which describes $N$ massless free scalar fields $\phi^i$ produces a second example of a non-analytic classical Lagrangian. The resulting deformed theory can also be obtained from the $4d$ Modified Maxwell theory by dimensional reduction \cite{Conti:2022egv}, so we will sometimes refer to this model as the Modified Scalar theory. This model is described by the action
\begin{align}\label{modified_scalar}
    S_{\text{Modified Scalar}} ( \gamma ) = \frac{1}{2} \int d^2 x \, &\Big( \cosh ( \gamma ) \partial_\mu \phi^i \partial^\mu \phi^i \nonumber \\
    &\quad + \sinh ( \gamma ) \sqrt{ 2 \partial_\mu \phi^i \partial_\nu \phi^i \partial^\nu \phi^j \partial^\mu \phi^j - \left( \partial_\mu \phi^i \partial^\mu \phi^i \right)^2 } \Big) \, ,
\end{align}
where the index $i = 1$, $\ldots$, $N$ labels the $N$ scalars. This action satisfies the flow equation
\begin{align}
    \frac{\partial S_{\text{Modified Scalar}} ( \gamma )}{\partial \gamma} = \mathcal{R}^{(2)} \, , 
\end{align}
as shown in \cite{Ferko:2022cix,Babaei-Aghbolagh:2022leo}. Like the $4d$ ModMax theory, the Modified Scalar theory is classically conformally invariant and thus the trace of its stress tensor vanishes, so only the term $T^{\mu \nu} T_{\mu \nu}$ appearing under the square root in $\mathcal{R}^{(2)}$ gives a non-zero contribution.

The two-dimensional root-$\TT$ deformation which gives rise to the theory (\ref{modified_scalar}) appears to share some of the interesting properties of the $\TT$ deformation, such as preserving classical integrability in several examples \cite{Borsato:2022tmu}. However, unlike the case of $\TT$, it is not yet known how to define the root-$\TT$ deformation at the quantum level and obtain flow equations for quantities like the S-matrix; proposed flow equations for the finite-volume spectrum and torus partition function of a root-$\TT$ deformed CFT were given in \cite{Ebert:2023tih} and supported using evidence from holography, but there is no general proof of these results. This presents an obstruction to carrying out the procedure that we have described above -- namely, identifying the correct prescription for the quantization of these models using some additional input -- for root-$\TT$ deformed models such as (\ref{modmax_lagrangian}) and (\ref{modified_scalar}).

In this work, we take up the task of studying the quantization of such non-analytic models in a simplified setting where one \emph{can} carry out this program explicitly, namely in the arena of $(0+1)$-dimensional theories. By performing a particular dimensional reduction described in \cite{Ferko:2023ozb}, either the ModMax theory or its Modified Scalar analogue can be reduced to a $1d$ model which describes a harmonic oscillator with a non-analytic interaction term. We refer to this system, which was first studied in \cite{Garcia:2022wad}, as the ModMax oscillator. The simplest version of this theory features two position variables $x(t)$, $y(t)$, and is described by the Lagrangian
\begin{align}\label{modmax_osc_lagrangian_intro}
    L_{\text{ModMax oscillator}} ( \gamma ) &= \frac{1}{2} \int dt \, \Bigg( \cosh ( \gamma )  \left( \dot{x}^2 + \dot{y}^2 - x^2 - y^2 \right)  \nonumber \\
    &\quad + \sinh ( \gamma ) \sqrt{ \left( \left( \dot{x} + y \right)^2 + \left( x - \dot{y} \right)^2 \right) \left( \left( \dot{x} - y \right)^2 + \left( x + \dot{y} \right)^2 \right) } \, \Bigg) \, .
\end{align}
When $\gamma = 0$, this theory reduces to a two-dimensional isotropic harmonic oscillator with unit mass and frequency. This gives the third example of a non-analytic theory.\footnote{There are several other examples of related non-analytic theories, which we will not discuss in detail here: a supersymmetric extension of ModMax \cite{Bandos:2021rqy,Kuzenko:2021cvx}, a $6d$ ModMax-like tensor theory \cite{Bandos:2020hgy}, and a $4d$ duality-invariant supersymmetric theory which is referred to as the MadMax sigma model \cite{Kuzenko:2023ysh}.}

At first glance, it is not obvious that the full Lagrangian (\ref{modmax_osc_lagrangian_intro}) at finite $\gamma$ will be amenable to exact quantization because of the velocity-dependent square root interaction. However, one might become more optimistic about the prospects of quantization after observing that this Lagrangian obeys a root-$\TT$-like flow equation,
\begin{align}\label{modmax_Blow}
    \frac{\partial L_{\text{ModMax oscillator}} ( \gamma )}{\partial \gamma} = \sqrt{ E_\gamma^2 - J_\gamma^2 } \, , 
\end{align}
where $E_\gamma$ and $J_\gamma$ are the energy and angular momentum, respectively, of the theory (\ref{modmax_osc_lagrangian_intro}) at parameter $\gamma$. This is the dimensional reduction of the flow equations driven by $\mathcal{R}^{(4)}$ and $\mathcal{R}^{(2)}$ that are obeyed by the ModMax and Modified Scalar models, respectively.

Because the ModMax oscillator can be described as a deformation of the harmonic oscillator by conserved charges, this suggests that one might perform canonical quantization of this theory using the prescription described in \cite{Gross:2019ach,Gross:2019uxi} for quantum mechanical deformations by functions of the Hamiltonian. That is, we first choose a basis of simultaneous eigenfunctions of the Hamiltonian and total angular momentum operators in the undeformed harmonic oscillator theory. We then declare that the eigenfunctions of the ModMax oscillator are the same as those of the harmonic oscillator, but with energy eigenvalues that have been shifted by the square root combination appearing in (\ref{modmax_Blow}). 

This prescription gives a simple and elegant way to define a quantum theory of the ModMax oscillator. However, as we have emphasized, quantization is not a functor: it is not clear that this is the only prescription, or even the correct one. For instance, because this quantization scheme is so simple, one might expect that it is also possible to quantize (\ref{modmax_osc_lagrangian_intro}) using the path integral formulation and get equivalent results. However, it is generally very difficult to perform the path integral for any theory which is not quadratic in derivatives. One of our goals in this work is to perform a detailed comparison of quantization prescriptions for the ModMax oscillator and check that they agree.

We will show that any deformation of a $1d$ theory by conserved charges induces a flow equation for the propagator, or partition function, of the theory. Remarkably, for the case of the deformation (\ref{modmax_Blow}), this flow equation is the Laplace equation in cylindrical coordinates. We will see that this flow equation for the partition function is consistent with, and can be derived from, either the canonical quantization prescription or the phase space path integral formulation. This strategy of reformulating the quantization of a deformed theory in terms of a flow equation for the partition function, or some other quantity, might be useful for cases where one cannot quantize the classical Lagrangian or Hamiltonian directly. This is the main motivation for the present work: we perform a detailed analysis of the quantization of this simple $1d$ model in the hope that some of the insights from studying this problem may be useful in understanding the quantization of other non-analytic models, such as the ModMax theory itself.\footnote{See \cite{Lechner:2022qhb} for a discussion of equivalent classical forms of ModMax which may be useful for quantization.}

The layout of this paper is as follows. In section \ref{sec:classical}, we study general classical deformations in a class of $1d$ theories, and prove basic results such as the equivalence of deformations in the Lagrangian and Hamiltonian formulations. In section \ref{sec:quantum}, we consider deformations of quantum mechanical theories by conserved charges using both canonical quantization and the path integral formalism, and derive flow equations for quantities like the propagator and partition function. In section \ref{sec:modmax_osc}, we apply the machinery developed in previous sections to the theory of the ModMax oscillator; understanding the quantum mechanical properties of this model is the main goal of the present work. In section \ref{sec:conclusion}, we summarize our results and identify some directions for future research. A first-order check of the equivalence of Lagrangian and Hamiltonian flows appears in appendix \ref{app:first_order}.

\section{Classical Deformations}\label{sec:classical}

In this section, we will consider deformations of $(0+1)$-dimensional theories at the classical level. We focus on theories that describe the dynamics of a collection of real bosons $x^i ( t ) $, $i = 1, \ldots, N$. The generalization to theories with fermions $\psi^i ( t )$ or more general degrees of freedom is straightforward, although we will not consider such cases here.\footnote{A convenient way to incorporate fermions is to define flow equations in superspace. These manifestly supersymmetric flows have been extensively studied; see \cite{Ferko:2021loo} and references therein for a review of such deformations in field theory, or \cite{Ebert:2022xfh,Ferko:2023ozb} for the corresponding flows in $1d$ theories.}

We view such a theory as being defined by either a Lagrangian or a Hamiltonian,
\begin{align}\label{L_and_H_def}
    L ( x^i , \dot{x}^i )  \; \text{ or } \; H ( x^i , p^i ) \, , 
\end{align}
where $p^i$ is the momentum which is canonically conjugate to the variable $x^i$,
\begin{align}
    p^i = \frac{\partial L}{\partial \dot{x}^i} \, .
\end{align}
Throughout this work, we assume that the indices $i$, $j$, etc. that label positions and momenta are raised or lowered with the trivial Euclidean metric $\delta_{ij}$. We therefore do not distinguish between upstairs and downstairs indices.

We will often write expressions like (\ref{L_and_H_def}) in which the argument of a function of positions, velocities, or momenta carries an index like $i$. In such expressions, the index $i$ is not meant to be a free index, but is merely shorthand to indicate dependence on all of the corresponding variables as $i$ runs from $1$ to $N$. Explicitly,
\begin{align}
    L ( x^i, \dot{x}^i ) = L \left( x^1 , \ldots , x^N , \dot{x}^1 , \ldots, \dot{x}^N \right) \, .
\end{align}

\subsection{Flow Equations for Lagrangians and Hamiltonians}

Our first goal is to study deformations, or flows, in the space of classical theories, which are defined as follows. Let $\mathcal{O} ( x^i , \dot{x}^i ; \lambda )$ be some function of the coordinates $x^i$ and their time derivatives, which may also depend on a real variable $\lambda$. The differential equation
\begin{align}\label{lagrangian_flow}
    \frac{\partial L}{\partial \lambda} = \mathcal{O} ( x^i , \dot{x}^i ; \lambda ) \, , 
\end{align}
defines a one-parameter family of Lagrangians $L ( x^i , \dot{x}^i ; \lambda )$. We refer to equation (\ref{lagrangian_flow}) as a flow equation and we say that the function $\mathcal{O}$ is the operator which drives the flow.\footnote{The use of the term ``operator'' is motivated by similar flow equations in field theories, such as the deformation of a $2d$ QFT by the $\TT$ operator. In the present context, $\cO$ is not a true quantum mechanical operator acting on a Hilbert space, but merely a classical function of positions and velocities.}

A theory may equivalently be described in the Hamiltonian formulation by the function
\begin{align}\label{hamiltonian_defn}
    H ( x^i , p^i ) = p^j \dot{x}^j - L ( x^i, \dot{x}^i ) \, , 
\end{align}
which is the Legendre transform of $L ( x^i, \dot{x}^i )$. In equation (\ref{hamiltonian_defn}), one must view all instances of the velocities $\dot{x}^i = \dot{x}^i ( p^j )$ as being implicitly defined in terms of the conjugate momenta.

In analogy with the Lagrangian flow (\ref{lagrangian_flow}), we may consider a differential equation
\begin{align}\label{hamiltonian_flow}
    \frac{\partial H}{\partial \lambda} = \fO \left( x^i , p^i ; \lambda \right) \, .
\end{align}
We use the symbol $\fO ( x^i , p^i )$, which is a function of positions and canonical momenta, to distinguish it from the function $\cO ( x^i , \dot{x}^i )$ that defines the flow equation for the Lagrangian.

Because we are interested in both Lagrangian and Hamiltonian descriptions of a physical system, it is natural to ask how the flow equations (\ref{lagrangian_flow}) and (\ref{hamiltonian_flow}) are related. A general deformation will modify both the Lagrangian $L$ or Hamiltonian $H$ and the relationship between the velocities $\dot{x}^i$ and the conjugate momentum $p^i$. Because such a deformation has two effects, one should check explicitly whether the diagram
\begin{equation}\label{commuting_diagram}
\begin{tikzcd}
	{L_0} && {L_\lambda} \\
	\\
	{H_0} & {} & {H_\lambda}
	\arrow["{\text{Deform by } \cO}", from=1-1, to=1-3]
	\arrow["{\substack{\text{Legendre} \\ \text{transform} } }"', from=1-1, to=3-1]
	\arrow["{\text{Deform by } \fO}"', from=3-1, to=3-3]
	\arrow["{\substack{\text{Legendre} \\ \text{transform} } }", from=1-3, to=3-3]
\end{tikzcd}
\end{equation}
commutes, and if so, under what conditions.

Fortunately, it turns out that the Lagrangian flow equation (\ref{lagrangian_flow}) and Hamiltonian flow equation (\ref{hamiltonian_flow}) lead to deformed quantities $L_\lambda$ and $H_\lambda$ which are related by a Legendre transform, as expected, so long as the operators $\cO$ and $\fO$ are related in the appropriate way. This result is stated more precisely in the following theorem.

\begin{theorem}\label{commute_theorem}

Let $L_0 ( x_i , \dot{x}_i )$ be a Lagrangian for a collection of coordinates $x_i ( t )$ and let $H_0 ( x_i , p_i )$ be the corresponding Hamiltonian. Given a function $\mathcal{O} \left( x^i , \dot{x}^i ; \lambda \right)$, consider a one-parameter family of Lagrangians $L_\lambda$ which satisfy the differential equations
\begin{align}\label{flow_thm}
    \frac{\partial L_\lambda}{\partial \lambda} = \mathcal{O} \left( x^i , \dot{x}^i ; \lambda \right) \, ,
\end{align}
with the initial conditions $L_\lambda \to L_0$ as $\lambda \to 0$. Then the Hamiltonian associated with $L_\lambda$,
\begin{align}
    H_\lambda ( x^i, p^i ) = p^j \dot{x}^j - L_\lambda ( x^i, \dot{x}^i ) \, , 
\end{align}
satisfies the flow equation
\begin{align}
    \frac{\partial H}{\partial \lambda} = \fO \left( x^i , p^i ; \lambda \right) \, ,
\end{align}
where the function $\fO$ is defined by
\begin{align}
    \fO \left( x^i , p^i ; \lambda \right) = - \cO \left( x^i , \dot{x}^i ( p^j ; \lambda  ) ; \lambda \right) \, , 
\end{align}
and where $\dot{x}^i ( p^j ; \lambda )$ represents the functional dependence between the velocities and conjugate momentum in the theory $L_\lambda$.

Conversely, given a function $\fO \left( x^i , p^i ; \lambda \right)$, consider the family of Hamiltonians $H_\lambda$ which obey
\begin{align}\label{hamiltonian_flow_theorem}
    \frac{\partial H_\lambda}{\partial \lambda} = \fO \left( x^i, p^i ; \lambda \right) \, ,
\end{align}
with initial condition $H_\lambda \to H_0$ as $\lambda \to 0$. The Lagrangians $L_\lambda$ associated with $H_\lambda$,
\begin{align}
    L_\lambda ( x^i, \dot{x}^i ; \lambda ) = p^j \dot{x}^j - H_\lambda ( x^i, \dot{x}^i ; \lambda ) \, , 
\end{align}
satisfy the flow equation (\ref{flow_thm}), where the operator $\cO$ is defined by
\begin{align}
    \cO \left( x^i , \dot{x}^i ; \lambda \right) = - \fO \left( x^i , p^i ( \dot{x}^j ; \lambda ) ; \lambda \right) \, , 
\end{align}
and where $p^i ( \dot{x}^j ; \lambda )$ represents the functional dependence between the conjugate momenta and velocities in the theory $H_\lambda$.

\end{theorem}

The interpretation of this theorem is that the diagram (\ref{commuting_diagram}) commutes, so long as the Lagrangian deformation $\cO$ and the Hamiltonian deformation $\fO$ are correctly related using the constraint between the velocities and conjugate momenta in the theory at finite $\lambda$. Note that this is not the same as using the relationship between the $p^i$ and $\dot{x}^i$ in the undeformed theories $L_0$ and $H_0$. Unsurprisingly, if one uses the relation between $\dot{x}^i$ and $p^i$ which is valid in the seed theories, the corresponding flows commute only to leading order in the deformation parameter $\lambda$. This was proven for field theories describing a single field $\phi$ and conjugate momentum $\pi = \frac{\partial \mathcal{L}}{\partial \dot{\phi}}$ in appendix A of \cite{Kruthoff:2020hsi}. To make the present work self-contained, we include the analogue of their leading-order proof for $(0+1)$-dimensional theories of $N$ positions $x^i ( t )$ in our appendix \ref{app:first_order}.

\begin{proof}

We first prove the forward direction, beginning with the flow equation (\ref{flow_thm}) for the Lagrangian. We view the velocities $\dot{x}^i$ as independent variables, while at each value of $\lambda$, the conjugate momenta $p^i ( \dot{x}^i ; \lambda )$ are determined via the relation
\begin{align}
    p^i ( \dot{x}^i ; \lambda ) = \frac{\partial L_\lambda}{\partial \dot{x}^i} \, .
\end{align}
The Legendre transform which defines the Hamiltonian, written in a way which emphasizes the $\lambda$ dependence, is
\begin{align}\label{legendre_proof_hamiltonian}
    H_\lambda \left( x^i , p^i ( \dot{x}^j ; \lambda ) \right) = p^j ( \dot{x}^k ; \lambda ) \dot{x}^j - L_\lambda ( x^i , \dot{x}^i ) \, .
\end{align}
We differentiate both sides of (\ref{legendre_proof_hamiltonian}) with respect to $\lambda$ to find
\begin{align}\label{proof_intermediate}
    \frac{\partial H_\lambda}{\partial p^i} \frac{\partial p^i}{\partial \lambda} + \frac{\partial H_\lambda}{\partial \lambda} = \frac{\partial p^j}{\partial \lambda} \dot{x}^j - \frac{\partial L_\lambda}{\partial \lambda} \, .
\end{align}
By the Hamilton equations of motion, we have
\begin{align}\label{hamilton_eom}
    \frac{\partial H_\lambda}{\partial p^i} = \dot{x}^i \, , 
\end{align}
and thus the terms $\frac{\partial p^j}{\partial \lambda} \dot{x}^j$ on both sides of equation (\ref{proof_intermediate}) cancel, leaving
\begin{align}\label{proof_intermediate_two}
    \frac{\partial H_\lambda}{\partial \lambda} = - \frac{\partial L_\lambda}{\partial \lambda} = - \cO ( x^i , \dot{x}^i ( p^j ; \lambda )  ; \lambda ) \, .
\end{align}
The object on the right side of (\ref{proof_intermediate_two}) is precisely the operator $\fO ( x^i , p^i ; \lambda )$. This completes the first half of the proof.

Now we show the reverse direction. Suppose that the Hamiltonian obeys the flow equation (\ref{hamiltonian_flow_theorem}). We now view the $p^i$ as independent variables while the $\dot{x}^i$ are fixed as
\begin{align}
    \dot{x}^i ( p^i ; \lambda ) = \frac{\partial H_\lambda}{\partial p^i} \, .
\end{align}
Again making all functional dependence explicit, the Legendre transform which defines the Lagrangian is
\begin{align}
    L_\lambda ( x^i , \dot{x}^i ( p^j ; \lambda ) ) = p^j \dot{x}^j ( p^i ; \lambda ) - H_\lambda ( x^i , p^i ) \, .
\end{align}
Differentiating with respect to $\lambda$ gives
\begin{align}\label{proof_reverse_intermediate}
    \frac{\partial L_\lambda}{\partial \dot{x}^i} \frac{\partial \dot{x}^i}{\partial \lambda} + \frac{\partial L_\lambda}{\partial \lambda} = p^j \frac{\partial \dot{x}^j}{\partial \lambda} - \frac{\partial H_\lambda}{\partial \lambda}  \, .
\end{align}
After using the relation
\begin{align}
    \frac{\partial L_\lambda}{\partial \dot{x}^i} = p^i \, , 
\end{align}
we see that the terms $p^j \frac{\partial \dot{x}^j}{\partial \lambda}$ cancel on either side of equation (\ref{proof_reverse_intermediate}), and we are left with
\begin{align}
    \frac{\partial L_\lambda}{\partial \lambda} = - \frac{\partial H_\lambda}{\partial \lambda} = - \fO \left( x^i , p^i ( \dot{x}^j ; \lambda ) ; \lambda \right) \, , 
\end{align}
which establishes the converse.
\end{proof}

Note that we have stated this theorem and its proof for $(0+1)$-dimensional theories. However, one can repeat this argument almost verbatim, making the replacements
\begin{align}
    x^i \to \phi^i \, , \quad p^i \to \pi^i \, , \quad L ( x^i, \dot{x}^i ) \to \mathcal{L} ( \phi^i, \dot{\phi}^i ) \, , \quad H ( x^i , p^i ) \to \mathcal{H} ( \phi^i, \pi^i ) \, , 
\end{align}
to obtain the corresponding theorem and its proof in any $d$-dimensional quantum field theory for a collection of fields $\phi^i$ and their conjugate momenta $\pi^i = \frac{\partial \mathcal{L}}{\partial \dot{\phi}^i}$.

\subsubsection*{\ul{\it Examples of Classical Lagrangian and Hamiltonian Deformations}}

It is instructive to see how Lagrangian and Hamiltonian flows are related in several examples, both when the assumptions of Theorem \ref{commute_theorem} hold and when they do not.

First let us consider a non-example of this theorem. We begin from an undeformed theory which has only a free kinetic term:
\begin{align}
    L_0 = \frac{1}{2} m \dot{x}^2 \, , \qquad H_0 = \frac{p^2}{2 m} \, .
\end{align}
The relationship between the undeformed velocity and momenta is simply $p = m \dot{x}$. Now consider the pair of flows
\begin{align}\label{simple_first_order_example}
    \frac{\partial L_\lambda}{\partial \lambda} = \cO ( x, \dot{x} ) = m \dot{x} \, , \qquad \frac{\partial H_\lambda}{\partial \lambda} = \fO ( x, p ) = - p \, ,
\end{align}
with solutions
\begin{align}
    L_\lambda = \frac{1}{2} m \dot{x}^2 + \lambda m \dot{x} \, , \qquad H_\lambda = \frac{p^2}{2 m} - \lambda p \, .
\end{align}
The Lagrangian $L_\lambda$ and Hamiltonian $H_\lambda$ are not related by a Legendre transformation. The conjugate momentum evaluated using the Lagrangian $L_\lambda$ is
\begin{align}
    p_\lambda = \frac{\partial L_\lambda}{\partial \dot{x}} = m \dot{x} + \lambda m \, , 
\end{align}
and thus the Legendre transform of $L_\lambda$, which we call $\widetilde{H}_\lambda$, is
\begin{align}
    \widetilde{H}_\lambda &= p_\lambda \dot{x} - L_\lambda = \frac{p^2}{2m} - \lambda p + \frac{1}{2} m \lambda^2 \, .
\end{align}
In this simple example, the difference between the Legendre transform $\widetilde{H}_\lambda$ and the Hamiltonian $H_\lambda$ is only a constant term, but nonetheless the two quantities do not agree. However, the difference $\widetilde{H}_\lambda - H_\lambda$ is of order $\lambda^2$. Thus $H_\lambda$ and $\widetilde{H}_\lambda$ agree to leading order in $\lambda$, which is consistent with the argument in appendix \ref{app:first_order}. In this example, the reason that the flows agree only at leading order is because the operators $\cO = m \dot{x}$ and $\fO = - p$ only satisfy the required constraint $\cO = - \fO$ if we use the relation between the velocity and conjugate momentum in the undeformed theories.

Next let us consider a modification of the above flow:
\begin{align}\label{momentum_flow}
    \frac{\partial L_\lambda}{\partial \lambda} = \cO ( x , \dot{x} ) = \frac{\partial L_\lambda}{\partial \dot{x}} \, , \qquad \frac{\partial H_\lambda}{\partial \lambda} = \fO ( x, p ) = - p \, .
\end{align}
This deformation agrees with (\ref{simple_first_order_example}) at leading order in $\lambda$, but beyond first order, it has been altered in order to satisfy the assumptions of Theorem \ref{commute_theorem}. We will solve these differential equations with initial conditions that are arbitrary functions of velocities or momenta, $L_0 ( \dot{x} ) $ and $H_0 ( p ) $. One finds
\begin{align}\label{simple_example_all_orders}
    L_\lambda = L_0 \left( \dot{x} + \lambda \right) \, , \qquad H_\lambda = H_0 - \lambda p \, .
\end{align}
The Lagrangian $L_\lambda$ and Hamiltonian $H_\lambda$ of (\ref{simple_example_all_orders}) are indeed related by a Legendre transform\footnote{This is an elementary property of the Legendre transform under translations. Let $f^\star ( p )$ represent the Legendre transform of a function $f(x)$. If $f ( x ) = g ( x + y )$, then $f^\star ( p ) = g^\star ( p ) - p y$.} to all orders in $\lambda$, as guaranteed by Theorem \ref{commute_theorem}.

Let us consider one more, slightly less trivial, example. Again beginning from an arbitrary velocity-dependent seed Lagrangian $L_0 ( \dot{x} )$ and Hamiltonian $H_0 ( p )$, consider the flow equations
\begin{align}\label{1d_TT_example}
    \frac{\partial L_\lambda}{\partial \lambda} = \left( L_\lambda - \dot{x} \frac{\partial L_\lambda}{\partial \dot{x}} \right)^2 \, , \qquad \frac{\partial H_\lambda}{\partial \lambda} = - H^2 \, .
\end{align}
This deformation was first considered in appendix A of \cite{Gross:2019ach}. The Hamiltonian flow equation can be solved for any initial condition $H_0 ( x, p )$:
\begin{align}\label{simple_hamiltonian_solution}
    H_\lambda = \frac{H_0}{1 + \lambda H_0} \, .
\end{align}
However, the Lagrangian flow equation is more complicated. If the seed theory is $L_0 = \dot{x}^2$, the solution is given in terms of a hypergeometric function:
\begin{align}
    L_\lambda = \frac{3}{4 \lambda} \left( {}_{3} F_2 \left[ - \frac{1}{2} , - \frac{1}{4} , \frac{1}{4} ; \frac{1}{3} , \frac{2}{3} ; \frac{256}{27} \lambda \dot{x}^2 \right] - 1 \right) \, .
\end{align}
This same hypergeometric function has appeared in several contexts related to classical deformations by conserved charges, including the $\TT$ deformation of the $2d$ Maxwell theory \cite{Conti:2018jho} and Yang-Mills \cite{Brennan:2019azg}. Despite the complicated form of $L_\lambda$, one can check that it is indeed related to the simpler function $H_\lambda$ of (\ref{simple_hamiltonian_solution}) by a Legendre transform when $H_0 = \frac{1}{4} p^2$. This is again required by the general argument of Theorem \ref{commute_theorem}.

\subsection{Deformations by Conserved Charges}\label{sec:classical_charge_deformations}

The two examples (\ref{momentum_flow}) and (\ref{1d_TT_example}) considered in the previous section are especially natural because they correspond to deformations of the theory by conserved quantities. Indeed, in equation (\ref{momentum_flow}) we deform the Lagrangian or Hamiltonian by the conjugate momentum $p$, which is conserved because the cyclic coordinate $x$ does not appear in the Lagrangians $L_\lambda ( \dot{x} )$. Likewise, equation (\ref{1d_TT_example}) is a deformation involving the Noether charge associated with time translation symmetry, which is the total energy of the system.

Such deformations by conserved charges are convenient to work with, since they give us a straightforward way to satisfy the conditions of Theorem \ref{commute_theorem}. This is because a conserved quantity, associated with a particular symmetry, can be easily characterized in either the Lagrangian or Hamiltonian formalism. In the former case we use Noether's theorem. If the Lagrangian is shifted by a total time derivative under the action of a symmetry generator $\delta$,
\begin{align}
    \delta L = \frac{df}{dt} \, , 
\end{align}
then the corresponding Noether charge
\begin{align}\label{noether_charge}
    Q = \frac{\partial L}{\partial \dot{x}^i} \delta x^i + f \, , 
\end{align}
obeys $\frac{d Q}{dt} = 0$ when the equations of motion are satisfied. Likewise, in the Hamiltonian formulation -- assuming that the charge $Q$ does not depend explicitly on time -- the relation
\begin{align}
    \frac{d Q}{dt} = \{ Q , H \} = 0 \, , 
\end{align}
holds on-shell.

Such conserved quantities typically have a clear physical interpretation, such as an energy or angular momentum, which makes them easy to describe either as functions of $x^i$ and $\dot{x}^i$ or as functions of $x^i$ and $p^i$. This is in contrast to a deformation of the Lagrangian by an arbitrary combination $\cO ( x^i , \dot{x}^i )$ of kinematical variables, for which one would have to explicitly work out the dependence $\dot{x}^i ( p^j )$ in order to find the corresponding Hamiltonian deformation $\fO ( x^i, p^i )$. Because of the naturalness of deformations by conserved charges, and their relationship to interesting higher-dimensional deformations like $\TT$ and root-$\TT$, we will focus on this class of flows in the remainder of this work.

More precisely, what we mean by a deformation by conserved charges is the following. Suppose that a seed theory $L_0$ has a collection of symmetries generated by variations $\delta_a$ and which are associated with conserved charges $Q_a$, for $a = 1, \ldots , M$, according to equation (\ref{noether_charge}). We will always use early Latin indices like $a, b, c$ to label charges $Q_a$ and middle Latin indices like $i, j, k$ to refer to coordinates $x^i$. We would like to define a flow equation of the form
\begin{align}\label{general_charge_flow}
    \frac{\partial L_\lambda}{\partial \lambda} = f \left( Q_1^{(\lambda)} ( x^i , \dot{x}^i ) , \ldots, Q_M^{(\lambda)} ( x^i , \dot{x}^i )  \right) \, ,
\end{align}
where $Q_a^{(\lambda)}$ is the Noether charge associated with the symmetry $\delta_a$ in the theory $L_\lambda$. The corresponding Hamiltonian flow is
\begin{align}\label{general_charge_flow_ham}
    \frac{\partial H_\lambda}{\partial \lambda} = - f \left( Q_1^{(\lambda)} ( x^i, p^i ), \ldots, Q_M^{(\lambda)} ( x^i, p^i ) \right) \, .
\end{align}
In this equation, the Hamiltonian charges $Q_a^{(\lambda)} ( x^i, p^i )$ are obtained by expressing the corresponding Lagrangian charges $Q_a^{(\lambda)} ( x^i , \dot{x}^i )$ in terms of conjugate momenta, using the relation between $p^i$ and $\dot{x}^i$ in the theory $L_\lambda$.

We must make an additional assumption in order to define such a flow: the variations $\delta_a$ must generate symmetries of the entire family of theories $L_\lambda$, rather than only for the seed theory $L_0$. That is, we must assume that this deformation does not break any of the symmetries. This will be true if all of the charges $Q_a$ are Poisson-commuting,
\begin{align}\label{poisson_commute}
    \{ Q_a, Q_b \} = 0 \, .
\end{align}
This assumption is sufficient because, if $Q_a$ is the Noether charge associated with a symmetry variation $\delta_a$, then the Poisson bracket with $Q_a$ generates the transformation of any function $g$ as $\{ Q_a, g \} = \delta_a g$. Thus the condition (\ref{poisson_commute}) implies that all of the charges $Q_a$ are invariant under all of the symmetries generating these charges, and thus any deformation of a Lagrangian by a function of these charges will still enjoy the same symmetries.

For instance, we could consider a theory with two coordinates $x^1 = x$ and $x^2 = y$ and which has two conserved momenta $p_x$, $p_y$ along with a conserved angular momentum $J$ and a Hamiltonian $H$. In this case,
\begin{align}
    \{ H , J \} = \{ H, p_x \} = \{ H , p_y \} = 0 \, , 
\end{align}
but one has
\begin{align}
    \{ J, p_x \} = p_y \, , \quad \{ J , p_y \} = - p_x \, .
\end{align}
In this case, we cannot define a flow equation (\ref{general_charge_flow}) if, for instance, the function $f$ depends on $J$ and $p_x$ but not $p_y$. Such a deformation breaks the rotational symmetry between $x$ and $y$, and thus $J$ is no longer a conserved quantity in the deformed theory. However, we are free to construct a flow equation using the three quantities
\begin{align}
    Q_1 = H \, , \quad Q_2 = J \, , \quad Q_3 = p_x^2 + p_y^2 \, ,
\end{align}
since
\begin{align}
    \{ J , p_x^2 + p_y^2 \} = 0 \, ,
\end{align}
and thus these three charges $Q_a$ Poisson-commute.

The first examples of deformations by conserved charges, which we will also call $f ( Q_a )$ flows, are those driven only by a function of the Hamiltonian $H$. This is the class of $f ( H )$ deformations considered in \cite{Gross:2019ach,Gross:2019uxi} and which includes the dimensional reduction of the $\TT$ deformation, in the special case where the $2d$ seed theory is conformally invariant. This dimensional reduction leads to the flow equation
\begin{align}\label{1d_TT_flow_defn}
    \partial_\lambda H_\lambda = \frac{H_\lambda^2}{\frac{1}{2} - 2 \lambda H_\lambda} \, , 
\end{align}
which has the solution
\begin{align}
    H_\lambda = \frac{1}{4 \lambda} \left( 1 - \sqrt{ 1 - 8 \lambda H_0 } \right) \, .
\end{align}
See \cite{Matsoukas-Roubeas:2022odk} for the extension of these $f(H)$ flows to the case of non-Hermitian Hamiltonian deformations. Another class of examples are those which involve the product of the conserved energy $E$ and a second charge $Q$, which were studied in \cite{Chakraborty:2020xwo} and which are similar to the $J \overbar{T}$ deformations of $2d$ field theories.

However, in the present work we will be primarily interested in the class of $f ( E, J^2 )$ deformations studied in \cite{Ferko:2023ozb}, which was motivated by the flow equation obeyed by the ModMax oscillator of \cite{Garcia:2022wad}. These deformations can be applied to theories which enjoy an additional $SO(N)$ symmetry which rotates the $N$ coordinates $x^i$ as
\begin{align}
    x^i ( t ) \longrightarrow \tensor{R}{^i_j} x^j ( t ) \, , \qquad R \in SO ( N ) \, .
\end{align}
The conserved currents associated with each of the rotation generators is a component of angular momentum,
\begin{align}
    J_{nm} &= \frac{\partial L}{\partial \dot{x}^n} x_m - \frac{\partial L}{\partial \dot{x}^m} x_n \, .
\end{align}
We define the total angular momentum by
\begin{align}
    J^2 = J^{nm} J_{nm} \, .
\end{align}
A general deformation by a function of both the energy and angular momentum, written in the Lagrangian formulation, is then
\begin{align}\label{general_f_H_Jsq}
    \frac{\partial L_\lambda}{\partial \lambda} = \cO \left( E , J^2 \right) \, ,
\end{align}
for some function $\cO$. We can, of course, write an equivalent flow equation for the Hamiltonian, $\partial_\lambda H_\lambda = \fO ( E, J^2 )$, where it is understood that $E$ and $J^2$ are functions of the positions $x^i$ and canonical momenta $p^i$ in the Hamiltonian flow.

The main operator of interest in the present work is
\begin{align}\label{rTT_1d}
    \mathcal{R} = \sqrt{ E ( x^i, \dot{x}^i ) ^2 - J ( x^i, \dot{x}^i ) ^2 } \, , \; \text{ or } \;  \mathfrak{R} = \sqrt{ H ( x^i, p^i ) ^2 - J ( x^i, p^i ) ^2 } \, , 
\end{align}
where, following the conventions for $\cO$ and $\fO$, we use calligraphic letters to refer to operators that depend on configuration space variables and Fraktur symbols for functions of phase space coordinates. In physically interesting examples, such as the $N$-dimensional harmonic oscillator, one always has the bound $| E |^2 \geq | J |^2$ so that the argument of the square root in (\ref{rTT_1d}) is non-negative; we will assume this to be true in what follows. We will always use the symbol $\gamma$ for the flow parameter when deforming by the operator $\mathcal{R}$, in contrast with $\lambda$, which we use as the parameter for a general deformation.

We refer to the combination (\ref{rTT_1d}), in either formulation, as the $1d$ root-$\TT$ operator. The reason for using this term is that, as shown in \cite{Ferko:2023ozb}, this object arises from a certain dimensional reduction of the flow equation
\begin{align}\label{root_TT_action_flow}
    \frac{\partial S_\gamma}{\partial \gamma} = \int d^2 x \, \sqrt{ \frac{1}{2}  T^{(\gamma) \mu \nu} T_{\mu \nu}^{(\gamma)} - \frac{1}{4} \left( \tensor{T}{^{(\gamma)}^\mu_\mu} \right)^2 } \, ,
\end{align}
which defines the root-$\TT$ deformation of $(1+1)$-dimensional field theories \cite{Ferko:2022cix}. More precisely, given a field theory describing the dynamics of a collection of scalar fields $\phi^i ( x, t )$ on a spatial circle $x \sim x + 2 \pi R$, one can Fourier-expand each scalar field as
\begin{align}\label{fourier_phi}
    \phi^i ( x, t ) = \sum_{n = - \infty}^{\infty} c^i_n ( t ) \exp \left( \frac{i n x}{R} \right) \, .
\end{align}
Truncating the theory to the dynamics of a single non-zero mode $c^i_m ( t )$, $m > 0$, and integrating over the circle then yields a dimensionally reduced theory for the functions $c^i_m ( t ) $. Performing this reduction for the family of theories that arises from deforming a collection of free scalars by the root-$\TT$ operator (\ref{root_TT_action_flow}) then yields a family of $1d$ theories which satisfy a flow equation driven by the combination (\ref{rTT_1d}). In particular, applying this $1d$ root-$\TT$ deformation to a seed theory of $N$ bosons $x^i ( t )$ subject to a harmonic oscillator potential yields the theory which we refer to as the ModMax oscillator. This will be the subject of section \ref{sec:modmax_osc}.

To conclude this subsection, we point out that -- although deformations by conserved charges are quite general -- not all models of interest satisfy flow equations of this form. Interesting non-examples include the Born oscillator, and generalized Born oscillator, which were recently studied in \cite{CoppaThesis,Giordano:2023byh}. A version of the Born oscillator for $N$ coordinates $x^i$ can be described by the Hamiltonian
\begin{align}
    H_\lambda = \frac{1}{\lambda} \left( \sqrt{ \left( 1 + \lambda p^i p^i \right) \left( 1 + \lambda x^i x^i \right) } - 1 \right) \, .
\end{align}
Despite its very symmetrical form, this Hamiltonian has the property that
\begin{align}
    \left\{ H_\lambda, \partial_\lambda H_\lambda \right\} \neq 0 \, , 
\end{align}
and thus it cannot obey any flow equation of the form $\partial_\lambda H_\lambda = \fO \left( H_\lambda, Q_a \right)$.

\subsection{Flow of the Classical Partition Function}\label{sec:classical_partition_function}

We are ultimately interested in quantum observables, such as the propagator, for theories deformed by functions of conserved charges, which will be studied in section \ref{sec:quantum}. The periodic Euclidean-time propagator reproduces the quantum thermal partition function, and the classical limit of this object is the ordinary classical partition function. It will therefore be useful to study the classical partition function in order to have a check against which to compare the results of section \ref{sec:quantum}, since these quantities should agree in the limit $\hbar \to 0$. We will see that the flow equations satisfied by the classical and quantum partition functions under a general deformation by conserved charges are identical.

Consider a theory with Hamiltonian $H ( x^i, p^i )$, for $i = 1$, $\ldots$, $N$, and a deformation by a function of conserved charges $Q_a$, $a = 1 , \ldots, M$, of the form in equation (\ref{general_charge_flow_ham}). One could also define $Q_0 = H$ to be the conserved charge associated with time translations, which would give $M+1$ charges in total. We define the grand canonical partition function
\begin{align}\label{grand_canonical_defn}
    \mathcal{Z} ( \beta, \lambda, \mu_a ) = \frac{1}{\left( 2 \pi \hbar \right)^N} \int d x^1 \, \ldots \, d x^N \, d p^1 \ldots d p^N \, \exp \left( - \beta H_\lambda ( x^i, p^i ) + \sum_{a = 1}^M \mu_a Q_a \right) \, .
\end{align}
Here $\beta = \frac{1}{T}$ is the inverse temperature and $\mu_a$ is the chemical potential, or fugacity, for the conserved charge $Q_a$. Note that the sum over charges in (\ref{grand_canonical_defn}) begins at $a = 1$ so that the Hamiltonian is not included, although we could treat the Hamiltonian symmetrically by defining $\mu_0 = - \beta$ and beginning the sum at $a = 0$. From now on, we will set $\hbar = 1$.

It is convenient to define the expectation value of a function $f$ in this ensemble as
\begin{align}
    \langle f ( x^j, p^j )  \rangle = \frac{1}{\left( 2 \pi \right)^N} \int d x^1 \, \ldots \, d x^N \, d p^1 \ldots d p^N \, f ( x^j, p^j )  \exp \left( - \beta H_\lambda ( x^i, p^i ) + \sum_{b = 1}^M \mu_b Q_b \right) \, .
\end{align}
The derivative of the partition function with respect to $\lambda$ is
\begin{align}
    \partial_\lambda \mathcal{Z} = - \beta \left\langle \partial_\lambda H_\lambda \right\rangle = - \beta \left\langle \fO ( H_\lambda , Q_a )  \right\rangle \, ,
\end{align}
and the derivatives with respect to the inverse temperature and chemical potentials are
\begin{align}
    \partial_\beta \mathcal{Z} = - \langle H_\lambda \rangle \, , \qquad \partial_{\mu_a} \mathcal{Z} = \langle Q_a \rangle \, .
\end{align}
The partition function therefore obeys the flow equation
\begin{align}\label{partition_function_flow}
    \partial_\lambda \mathcal{Z} = - \beta \fO \left[ - \partial_\beta , \partial_{\mu_1} , \ldots , \partial_{\mu_M} \right] \left(  \mathcal{Z} \right) \, .
\end{align}
The right side of (\ref{partition_function_flow}) is defined by expanding the operator $\fO$ as a power series in each of its variables $H_\lambda$, $Q_a$, and then replacing each variable with the appropriate derivative. This is only possible if the deforming operator is an analytic function of its arguments, but we will see shortly how to extend this argument to some non-analytic deformations.

For example, let us consider the flow equation for the one-dimensional $\TT$ deformation,
\begin{align}\label{1d_TT_series_Z}
    \partial_\lambda H_\lambda = \frac{H_\lambda^2}{\frac{1}{2} - 2 \lambda H_\lambda} = 2 H_\lambda^2 + 8 \lambda H_\lambda^3 + 32 \lambda^2 H_\lambda^4 + 128 \lambda^3 H_\lambda^5 + \ldots \, .
\end{align}
We may schematically write the corresponding flow equation for the partition function as
\begin{align}
    \partial_\lambda \mathcal{Z} = - \beta \left( \frac{\partial_\beta^2}{\frac{1}{2} + 2 \lambda \partial_\beta } \right) \mathcal{Z} \, , 
\end{align}
where the rational function of derivatives is again defined by the Taylor series expansion whose first few terms are shown in (\ref{1d_TT_series_Z}). We can invert this infinite series of derivatives to write the equivalent flow equation
\begin{align}
    \left( \frac{1}{2} + 2 \lambda \partial_\beta \right) \left( \frac{1}{\beta} \partial_\lambda \mathcal{Z} \right) = - \partial_\beta^2 \mathcal{Z} \, , 
\end{align}
which can be expressed as
%
%
\begin{align}\label{TT_QM_partition_function_flow}
    \left( 4 \lambda \partial_\lambda \partial_\beta + 2 \beta \partial_\beta^2 + \left( 1 - \frac{4 \lambda}{\beta} \right) \partial_\lambda \right) \mathcal{Z} ( \lambda , \beta ) = 0 \, .
\end{align}
This is the flow equation for the classical partition function of a theory deformed by the $1d$ $\TT$ flow. It also turns out that the quantum partition function obeys the same flow equation.\footnote{For the negative sign of the deformation parameter, this also matches the differential equation one obtains from studying JT gravity with a finite radial cutoff \cite{Iliesiu:2020zld}.} We will see in section \ref{sec:quantum} that this is true more generally: the flow equations for the classical and quantum partition functions are always identical for deformations by any function of conserved charges.

An interesting feature of the differential equation (\ref{TT_QM_partition_function_flow}) is that its solution can be written as an integral transform of the undeformed partition function $Z_0 ( \beta )$ at $\lambda = 0$. This can be understood as an analogue of the solution to the heat equation, written as a convolution against the heat kernel, since (\ref{TT_QM_partition_function_flow}) takes the form of a diffusion-type equation where the parameter $\lambda$ plays the role of time. This integral kernel solution is
\begin{align}\label{1d_TT_Z_integral_kernel}
    Z_\lambda ( \beta ) = \int_{0}^{\infty} d \beta' \, \frac{\beta}{\sqrt{- 8 \pi  \lambda } \beta^{\prime 3/2}} \exp \left( \frac{ ( \beta - \beta' )^2}{8 \lambda \beta'} \right) Z_0 ( \beta' ) \, ,
\end{align}
which was obtained and studied in \cite{Gross:2019uxi}. It is easy to check that the integral expression (\ref{1d_TT_Z_integral_kernel}) automatically solves the differential equation (\ref{TT_QM_partition_function_flow})

We now turn our attention to deformations which take the form
\begin{align}\label{square_root_Blow}
    \frac{\partial H_\lambda}{\partial \lambda} = \sqrt{ f ( H_\lambda, Q_a ) } \, , 
\end{align}
where $f$ is an analytic function. In the case where $f ( H_\lambda = 0 , Q_a = 0 ) = 0$, this deforming operator does not admit a Taylor series expansion because the square root function is not analytic around $0$. For such deformations, we cannot define the differential operator appearing on the right side of equation (\ref{partition_function_flow}) via a series.

We can attempt to circumvent this difficulty in one of two ways. The first way is to attempt to define a fractional derivative by diagonalizing the differential operator. That is, if we can identify a complete basis of eigenfunctions $\psi_n$ with the property
\begin{align}
    f \left( - \partial_\beta , \partial_{\mu_a} \right) \psi_n = \nu_n \psi_n \, , 
\end{align}
for some non-negative eigenvalues $\nu_n$, then we simply define the fractional differential operator to act as
\begin{align}
    \sqrt{ f \left( - \partial_\beta , \partial_{\mu_a} \right) } \psi_n = \sqrt{ \nu_n } \psi_n \, .
\end{align}
We may then expand the partition function $\mathcal{Z}$ in this basis as
\begin{align}
    \mathcal{Z} = \sum_{n} c_n \psi_n 
\end{align}
and obtain a flow equation
\begin{align}
    \partial_\lambda \mathcal{Z} = - \beta \sum_n \sqrt{ \nu_n } c_n \psi_n \, .
\end{align}
Although it should be possible, in principle, to carry out this procedure of defining the fractional derivative -- at least in some examples -- we will not pursue it further here.

Instead, we will attempt to remedy the non-analyticity by taking a second derivative. In certain cases, this can convert a first-order flow equation driven by a square-root operator to a second-order flow equation driven by an analytic operator. For a general deformation by an operator $\fO = \fO ( H_\lambda, Q_a )$, the second derivative of the partition function with respect to $\lambda$ is
\begin{align}
    \partial_\lambda^2 \mathcal{Z} = \left\langle - \beta \frac{\partial \fO}{\partial \lambda} + \beta^2 \fO^2 \right\rangle \, ,
\end{align}
and for an operator $\fO = \sqrt{ f ( H_\lambda, Q_a ) }$ of the form in equation (\ref{square_root_Blow}), this is
\begin{align}\label{second_order_Z_flow_sqrt}
    \partial_\lambda^2 \mathcal{Z} = \left\langle \frac{- \beta }{2}  \frac{\partial_\lambda f}{\sqrt{f}} + \beta^2 f \right\rangle \, .
\end{align}
The second term now depends only on $f$ but not its square root, so this term can be expressed in terms of a power series in derivatives of $\mathcal{Z}$ with respect to $\beta$ and the $\mu_a$ as before. The first term, however, still depends on $\sqrt{ f }$ and is not manifestly analytic for a deformation involving a generic function $f$. However, we will revisit this expression in section \ref{sec:modmax_osc} for the special case $f ( H_\lambda, Q_a ) = \sqrt{ H_\lambda^2 - J_\lambda^2 }$ and the seed theory of a harmonic oscillator. In this case, we will see that the first term also becomes analytic, and the flow equation for $\mathcal{Z}$ collapses to a conventional second-order differential equation in three variables. Although this PDE does not admit an integral kernel solution like (\ref{1d_TT_Z_integral_kernel}), its general solution can be written in terms of exponentials and Bessel functions.

\section{Quantum Deformations}\label{sec:quantum}

In this section we turn to the deformation of quantum-mechanical theories by conserved charges in $(0+1)$ spacetime dimensions. As we mentioned in the introduction, there is no universal method for quantizing a general classical theory; see, for instance, \cite{Ali:2004ft} for a survey of quantization methods from a mathematical perspective.

Here we will focus on canonical quantization and path integral quantization. When discussing operators in the canonical formalism, we will use hats to distinguish them from the corresponding classical variables; for instance, we write
\begin{align}
    \hat{x}^i \ket{ \vec{x} } = x^i \ket{ \vec{x} } \, , \qquad \hat{p}^i \ket{\vec{p}} = p^i \ket{\vec{p}} \, , \quad \text{etc} \, .
\end{align}
We will use vector symbols $\vec{x}$, $\vec{p}$ to represent the collection of all components $x^i$ and $p^i$, respectively, for $i = 1 , \ldots, N$. The component indices $i$, $j$, etc. are not to be confused with the subscripts $A$, $B$, which we will introduce shortly and which refer to the initial and final configurations that determine the boundary conditions of the path integral.

One of the results of section \ref{sec:classical} is that deformations by conserved charges in the Hamiltonian and Lagrangian formulations are equivalent, if one uses the correct relationship between variables in the deforming operators. Because the quantum theory features either the Hamiltonian, in canonical quantization, or the Lagrangian, in conventional path integral quantization, a natural first question to ask is whether deformations by conserved charges in these two formalisms are again equivalent quantum-mechanically.

For example, one might ask whether computing the propagator using the unitary time evolution operator $\widehat{U} ( t_B, t_A )$ associated with a given Hamiltonian $\widehat{H} ( t )$,
\begin{align}\label{ham_prop}
    K ( \vec{x}_B, t_B ; \vec{x}_A, t_A ) &= \langle \vec{x}_B \mid \widehat{U} ( t_B, t_A ) \mid \vec{x}_A \rangle \, , \nonumber \\
    \frac{\partial \widehat{U} ( t, t' )}{\partial t} \Big\vert_{t = t'} &= - i \widehat{H} ( t ) \, , 
\end{align}
agrees with a definition using the Feynman path integral,
\begin{align}\label{lag_prop}
    K ( \vec{x}_B, t_B ; \vec{x}_A, t_A ) = \int_{\vec{x} ( t_A ) = \vec{x}_A}^{\vec{x} ( t_B ) = \vec{x}_B } \mathcal{D} \vec{x} \, \exp \left( i \int_{t_A}^{t_B} dt \, L_0 \right) \, ,
\end{align}
after deforming $\widehat{H}_0$ by an operator $\widehat{\fO} ( \widehat{Q}_a ) $ in (\ref{ham_prop}) and deforming $L_0$ by the corresponding operator $\cO ( Q_a ) = - \fO ( Q_a )$ in (\ref{lag_prop}).

However, as we will review, this is not the right question to ask. The expression (\ref{lag_prop}) for the propagator, in terms of a path integral over the position coordinates $\vec{x}$ with the standard measure, is only valid for Lagrangians which are quadratic in derivatives. After deforming a seed theory by a function of conserved charges, the resulting deformed Lagrangian will often have more general dependence on the derivatives $\dot{x}^i$. In such situations, the conventional path integral (\ref{lag_prop}) does not correctly compute the deformed propagator, and instead one must use a phase space path integral:
\begin{align}\label{phase_space_path_integral}
    K ( \vec{x}_B, t_B ; \vec{x}_A, t_A ) = \int_{\vec{x} ( t_A ) = \vec{x}_A}^{\vec{x} ( t_B ) = \vec{x}_B} \, \mathcal{D} \vec{x} \, \int_{\vec{p} ( t_A ) = \vec{p}_A}^{\vec{p} ( t_B ) = \vec{p}_B } \, \mathcal{D} \vec{p} \, \exp \left( i \int_{t_A}^{t_B} dt \, \left( p^i ( t ) \dot{x}_i ( t ) - H_0 ( \vec{p}, \vec{x} ) \right) \right) \, ,
\end{align}
The quantity appearing in the exponential of (\ref{phase_space_path_integral}) is \emph{not} the classical action, because the functions $p^i ( t )$ in the first term are not the canonical momenta, but rather dummy functions which are path-integrated over. Thus, for a general non-quadratic action, the classical Lagrangian plays no role in the path integral quantization of the theory, and only the phase space path integral defined in (\ref{phase_space_path_integral}) is important.

Given this observation, we should not ask whether deformations of the Hamiltonian and Lagrangian are equivalent in the quantum theory. Instead we should ask whether deforming the Hamiltonian operator in the expression (\ref{ham_prop}) is equivalent to deforming the classical function of phase space variables $H_0 ( \vec{p}, \vec{x} )$ in (\ref{phase_space_path_integral}). This will be the topic of section \ref{sec:canonical_and_path_integral}. First we will take a detour to review the phase space path integral.

\subsection{Phase Space Path Integral}\label{sec:phase_space_path_integral}

In this section, we will review the path integral quantization of a general Hamiltonian 
\begin{align}
    \widehat{H} ( \hat{x}^i, \hat{p}^i ) \, ,
\end{align}
which need not be quadratic in the momenta $\hat{p}^i$ that are conjugate to the position operators $\hat{x}^i$. Only in the case of this quadratic dependence on momenta does the general phase space path integral reduce to the ordinary Feynman path integral.\footnote{This is emphasized in some, but not all, textbooks. See, for instance, Section 2.2 of \cite{Ramond:1981pw}, Section 10.2 of \cite{ZinnJustin2005PathII} or Section 1.2 of \cite{Mosel:2004mk}.}

To begin, we will assume that the Hamiltonian is an analytic function of the variables $\hat{x}^i$ and $\hat{p}^i$. Later we will be interested in deformations of such Hamiltonians by non-analytic functions of charges, but for now we will require $\widehat{H}$ to admit an expansion
\begin{align}\label{normal_ordered}
    \widehat{H} ( t ; \vec{x} , \vec{p} )  = \sum_{i, j, m, n} h_{i j m n} ( t ) \left( \hat{p}^i \right)^m \left( \hat{x}^j \right)^n \, .
\end{align}
One can always bring a general analytic Hamiltonian into this form by using the canonical commutation relation to move all position operators $\hat{x}^i$ to the right of momentum operators $\hat{p}^i$. The form (\ref{normal_ordered}) of the Hamiltonian is said to be normal-ordered. Taking the Hermitian conjugate of this expression reverses the order of the operators,
\begin{align}\label{anti_normal_ordered}
    \widehat{H}^\dagger ( t ; \vec{x} , \vec{p} ) = \sum_{i, j, m, n} h_{i j m n}^\ast ( t )  \left( \hat{x}^j \right)^n \left( \hat{p}^i \right)^m \, , 
\end{align}
and is thus anti-normal-ordered. We also assume that the Hamiltonian is Hermitian, so that $\widehat{H} = \widehat{H}^\dagger$. Our conventions for the position and momentum eigenstates are
\begin{align}
    \langle \vec{x} \mid \vec{p} \rangle = \exp \left( i x^i p_i \right) \, , 
\end{align}
and we take $\hbar = 1$. 

We are interested in computing the propagator
\begin{align}
    K \left( \vec{x}_B, t_B ; \vec{x}_A , t_A \right) =  \left\langle \vec{x}_B \mid \widehat{U} ( t_B, t_A ) \mid \vec{x}_A \right\rangle \, .
\end{align}
where the unitary time evolution operator $\widehat{U}$ is related to the Hamiltonian according to (\ref{ham_prop}). Following the usual time-slicing procedure, we subdivide the time interval $T = t_B - t_A$ into a large number $M + 1$ of smaller intervals of length $\epsilon$,
\begin{align}
    t_0 = t_A \, , \quad t_1 = t_A + \epsilon \, , \quad t_2 = t_A + 2 \epsilon \, , \quad \ldots \, , \quad t_{M} = t_A + M \epsilon \, , \quad t_{M+1} = t_B \, ,
\end{align}
where we have defined
\begin{align}
    \epsilon = \frac{T}{M + 1} \, .
\end{align}
The time evolution operator decomposes into a product of operators $\widehat{U} ( t_{j+1}, t_j )$ over each of the smaller time intervals,
\begin{align}
    \widehat{U} ( t_B, t_A ) = \prod_{j = 0}^{M} \widehat{U} ( t_{j+1} , t_j ) \, .
\end{align}
Furthermore, we can use the completeness of the position eigenstates
\begin{align}
    \int d \vec{x} \, \ket{\vec{x}} \bra{\vec{x}} = 1 \, , 
\end{align}
where the integration measure $d \vec{x}$ is shorthand for $\prod_{i=1}^{N} d x^i$, to insert several resolutions of the identity and write
\begin{align}\label{time_sliced_propagator}
    K \left( \vec{x}_B, t_B ; \vec{x}_A , t_A \right) = \int \left( \prod_{k = 1}^{M} d \vec{x}_k \right) \, \left( \prod_{j=0}^{M} \langle \vec{x}_{j+1} \mid \widehat{U} ( t_{j+1}, t_{j} ) \mid \vec{x}_{j} \rangle \right) \, .
\end{align}
Here we have defined $\vec{x}_0 = \vec{x}_A$ and $\vec{x}_M = \vec{x}_B$. Note that the subscripts $j$, $k$, etc. on the position variables do not refer to the components $x^i$ of the vector $\vec{x}$ but rather to labels which index the different integration variables.

We can now focus on the propagator over one of the smaller time intervals of length $\epsilon$. Over such an interval from $t$ to $t + \epsilon$, even if the Hamiltonian $\widehat{H}$ has explicit time dependence, we can approximate the unitary time evolution operator to $\mathcal{O} ( \epsilon )$ as
\begin{align}
    \widehat{U} ( t + \epsilon , t ) = e^{- i \widehat{H} ( t ) \epsilon} + \mathcal{O} ( \epsilon^2 ) \, .
\end{align}
Inserting a complete set of momentum eigenstates using the completeness relation
\begin{align}
    \int d \vec{p} \, \, \ket{\vec{p}} \bra{\vec{p}} = \frac{1}{( 2 \pi )^N} \, , 
\end{align}
we can then write a single factor in the integrand of (\ref{time_sliced_propagator}) as
\begin{align}\label{U_symmetric}
    \langle \vec{x}_{j+1} \mid U ( t + \epsilon, t ) \mid \vec{x}_j \rangle = \int \frac{d \vec{p}}{( 2 \pi )^N} \left\langle \vec{x}_{j+1} \mid \widehat{U} \left( t + \epsilon , t + \frac{\epsilon}{2} \right) \mid \vec{p} \right\rangle \left\langle \vec{p} \mid \widehat{U} \left( t + \frac{\epsilon}{2} , t \right) \mid \vec{x}_j \right\rangle \, . 
\end{align}
It is convenient that we have two representations of the Hamiltonian (\ref{normal_ordered}) and (\ref{anti_normal_ordered}), one with position operators to the right and one with momentum operators to the right. We evaluate the second matrix element in (\ref{U_symmetric}) to order $\epsilon$ using the normal-ordered form,
\begin{align}\label{second_element}
    \left\langle \vec{p} \mid \widehat{U} \left( t + \frac{\epsilon}{2} , t \right) \mid \vec{x}_j \right\rangle &= \left\langle \vec{p} \mid 1 - \frac{i \epsilon}{2} \widehat{H} ( t )  \mid \vec{x}_j \right\rangle + \mathcal{O} ( \epsilon^2 ) \nonumber \\
    &= \langle \vec{p} \mid \vec{x}_j \rangle - \frac{i \epsilon}{2} \left\langle \vec{p} \; \; \Big\vert \; \sum h_{ikmn} ( t )  \left( \hat{p}^i \right)^m \left( \hat{x}^k \right)^n \; \Big\vert \; \vec{x}_j \right\rangle + \mathcal{O} ( \epsilon^2 ) \nonumber \\
    &= e^{- i \vec{p} \cdot \vec{x}_j } - \frac{i \epsilon}{2} h ( t; \vec{x}_j , \vec{p} ) + \mathcal{O} ( \epsilon^2 ) \, , 
\end{align}
where we use the symbol $h ( \vec{x}_j , \vec{p} )$ with no indices to refer to the normal-ordered Hamiltonian (\ref{normal_ordered}) with all operators replaced with classical variables. 

Similarly, we use hermiticity of $\widehat{H}$ along with the anti-normal-ordered form (\ref{anti_normal_ordered}) for the Hamiltonian to evaluate the first matrix element appearing in (\ref{U_symmetric}),
\begin{align}\label{first_element}
    &\left\langle \vec{x}_{j+1} \mid \widehat{U} \left( t + \epsilon , t + \frac{\epsilon}{2} \right) \mid \vec{p} \right\rangle \nonumber \\
    &\quad = \left\langle \vec{p} \mid 1 - \frac{i \epsilon}{2} \widehat{H}^\dagger \left( t + \frac{\epsilon}{2} \right)  \mid \vec{x}_j \right\rangle + \mathcal{O} ( \epsilon^2 ) \nonumber \\
    &\quad = \langle \vec{x}_{j+1} \mid \vec{p} \, \rangle - \frac{i \epsilon}{2} \left\langle \vec{x}_{j+1} \; \; \Big\vert \; \sum h_{ikmn}^\ast \left( t + \frac{\epsilon}{2} \right)  \left( \hat{x}^k \right)^n \left( \hat{p}^i \right)^m  \; \Big\vert \; \vec{p} \right\rangle + \mathcal{O} ( \epsilon^2 ) \nonumber \\
    &\quad = e^{i \vec{x}_{j+1} \cdot \vec{p}} - \frac{i \epsilon}{2} h^\ast \left( t + \frac{\epsilon}{2} ; \vec{x}_{j+1} , \vec{p} \right) + \mathcal{O} ( \epsilon^2 ) \, , 
\end{align}
where we similarly write $h^\ast ( t; \vec{x}_{j+1} , \vec{p} )$ for the anti-normal-ordered Hamiltonian (\ref{normal_ordered}) with operators replaced by classical variables.

Using these results (\ref{second_element}) and (\ref{first_element}) for the matrix elements, we find
\begin{align}
    &\langle \vec{x}_{j+1} \mid U ( t + \epsilon, t ) \mid \vec{x}_j \rangle \nonumber \\
    &\quad = \int \frac{d \vec{p}}{ ( 2 \pi )^N} \left( e^{i \vec{x}_{j+1} \cdot \vec{p}} - \frac{i \epsilon}{2} h^\ast \left( t + \frac{\epsilon}{2} ;  \vec{x}_{j+1} , \vec{p} \right) \right) \left( e^{- i \vec{p} \cdot \vec{x}_j } - \frac{i \epsilon}{2} h ( t ; \vec{x}_j , \vec{p} ) \right) + \mathcal{O} ( \epsilon^2 ) \nonumber \\
    &\quad = \int \frac{d \vec{p}}{ ( 2 \pi )^N} e^{i ( \vec{x}_{j+1} - \vec{x}_j ) \cdot \vec{p}} \exp \left[ - \frac{i \epsilon}{2} \left( h \left(  t + \frac{\epsilon}{2} ; \vec{x}_j , \vec{p} \right) + h^\ast \left( t + \frac{\epsilon}{2} ; \vec{x}_{j+1} , \vec{p} \right) \right) \right]  + \mathcal{O} ( \epsilon^2 ) \, .
\end{align}
This discretization suggests that we should define the classical Hamiltonian
\begin{align}
    H ( t ;  \vec{x} , \vec{p} ) = \frac{1}{2} \left( h ( t ;  \vec{x} , \vec{p} ) + h^\ast ( t ;  \vec{x} , \vec{p} ) \right) = \Re \left( h ( t ;  \vec{x} , \vec{p} ) \right) \, .
\end{align}
To leading order at small $\epsilon$, the propagator over a small time interval $\epsilon$ is therefore
\begin{align}\label{smol_propagator}
    \langle \vec{x}_{j+1} \mid U ( t + \epsilon, t ) \mid \vec{x}_j \rangle = \int \frac{d \vec{p}}{( 2 \pi )^N} \exp \left[ i \left( \vec{p}_j \cdot \left( \vec{x}_{j+1} - \vec{x}_j \right) - \epsilon H ( t ; \vec{x}_j , \vec{p}_j ) \right) \right] \, .
\end{align}
The full time-sliced propagator (\ref{time_sliced_propagator}) is obtained from the product of the individual factors (\ref{smol_propagator}) in the limit as $\epsilon \to 0$ and $M \to \infty$,
\begin{align}\label{M_to_infty_propagator}
    &K \left( \vec{x}_B, t_B ; \vec{x}_A , t_A \right) \nonumber \\
    &\quad = \lim_{\epsilon \to 0} \int \left( \prod_{k=1}^{M} d \vec{x}_k \right) \int \left( \prod_{l = 0}^{M} \frac{d \vec{p}_l}{(2 \pi)^N} \right) \exp \left[ i  \sum_{j=0}^{M} \left( \vec{p}_j \cdot \left( \vec{x}_{j+1} - \vec{x}_j \right) - \epsilon H ( t ; \vec{x}_j , \vec{p}_j ) \right) \right] \, .
\end{align}
In the limit of small $\epsilon$, the sum in the argument of the exponential becomes an integral:
\begin{align}
    \lim_{\epsilon \to 0} \sum_{j=0}^{M} \epsilon \left( \vec{p}_j \cdot \frac{\left( \vec{x}_{j+1} - \vec{x}_j \right)}{\epsilon} - 
 H ( t ; \vec{x}_j , \vec{p}_j ) \right) = \int_{t_A}^{t_B} d t \,\left( p^i ( t ) \dot{x}_i ( t ) - H ( t ; \vec{x} ( t ) , \vec{p} ( t ) \right) \, .
\end{align}
Here we have passed from discrete collections of $\vec{x}_j$, $\vec{p}_j$ to continuous trajectories $x^i ( t )$, $p^i ( t )$. We conclude that the propagator for a general analytic Hamiltonian takes the form
\begin{align}\label{final_phase_space_path_integral}
    K ( \vec{x}_B, t_B ; \vec{x}_A, t_A ) = \int_{\vec{x} ( t_A ) = \vec{x}_A}^{\vec{x} ( t_B ) = \vec{x}_B} \, \mathcal{D} \vec{x} \, \int_{\vec{p} ( t_A ) = \vec{p}_A}^{\vec{p} ( t_B ) = \vec{p}_B } \, \mathcal{D} \vec{p} \, \exp \left( i \int_{t_A}^{t_B} dt \, \left( p^i ( t ) \dot{x}_i ( t ) - H ( \vec{p}, \vec{x} ) \right) \right) \, ,
\end{align}
where the path integral measures $\mathcal{D} \vec{x}$ and $\mathcal{D} \vec{p}$ are defined as the limits of the products in (\ref{M_to_infty_propagator}). To respect causality, we set the propagator equal to (\ref{final_phase_space_path_integral}) when $t_B > t_A$ and set $K ( \vec{x}_B, t_B ; \vec{x}_A, t_A ) = 0$ for $t_B < t_A$. Similarly, by performing this phase space path integral in Euclidean time with periodic boundary conditions, one can obtain a phase space integral expression for the finite-temperature partition function.

Again, it is important to emphasize that the integral (\ref{final_phase_space_path_integral}) runs over all phase space paths $(\vec{x} ( t ) , \vec{p} ( t ))$. For a generic path, there is no relationship between the coordinates and momenta; in particular, it is not the case that $p^i$ is constrained to be equal to the canonical momentum which is conjugate to $x^i$. In the special case where the Hamiltonian is quadratic in the momenta, for instance if
\begin{align}
    H = \frac{p^i p_i}{2m} + V ( x ) \, , 
\end{align}
then the path integral over momenta can be evaluated, leaving a path integral over positions $x^i ( t )$. Only when performing this evaluation is the expression $p^i ( t )$ set equal to the conjugate momentum, which reproduces the usual Feynman path integral (\ref{lag_prop}) which involves the Lagrangian. For non-quadratic Hamiltonians, however, no such reduction is possible and we must use the more fundamental form (\ref{final_phase_space_path_integral}).

\subsection{Deformations in Canonical Quantization and Path Integral Quantization}\label{sec:canonical_and_path_integral}

We now wish to study how observables in quantum mechanics are modified when the theory is deformed by conserved charges. Our goal is to show that the propagator, and hence the finite-temperature partition function, of a general theory satisfies a flow equation which is identical to the one derived in section (\ref{sec:classical_partition_function}) for the classical partition function. As a consistency check, we will also see that this flow equation for the propagator can be equivalently derived using either canonical methods or path integral methods.

\subsubsection*{\ul{\it Canonical Analysis}}

First let us see how to understand this flow equation in the canonical formalism. For the moment, we will specialize to the case of Hamiltonians which do not depend on time explicitly. In this case, the unitary time evolution operator can be written as
\begin{align}
    \widehat{U} ( t_B , t_A ) = e^{- i \widehat{H} ( t_B - t_A ) } \, ,
\end{align}
and the propagator is
\begin{align}\label{regular_propagator}
    K ( \vec{x}_B, t_B ; \vec{x}_A, t_A ) &= \langle \vec{x}_B \mid e^{- i \widehat{H} ( t_B - t_A ) } \mid \vec{x}_A \rangle \, .
\end{align}
Suppose that the one-parameter family of Hamiltonian operators $\widehat{H}_\lambda$ satisfies a flow equation driven by a combination of conserved charge operators $\widehat{Q}_a$,
\begin{align}
    \frac{\partial \widehat{H}_\lambda}{\partial \lambda} = \widehat{\fO} \left( \widehat{Q}_a \right) \, .
\end{align}
If the operator $\widehat{\fO}$ depends on charges besides the Hamiltonian, it will not be possible to derive a closed flow equation for the usual propagator (\ref{regular_propagator}). Instead, we must consider a more general propagator which includes sources for the various conserved charges $Q_a$. We therefore define the quantity
\begin{align}\label{flavored_propagator}
    K ( \vec{x}_B, t_B ; \vec{x}_A, t_A ; \lambda ; \mu_a ) &= \left\langle \vec{x}_B \; \Big\vert \; \exp \left( - i \widehat{H}_\lambda ( t_B - t_A ) + \sum_a \mu_a \widehat{Q}_a \right) \; \Big\vert \; \vec{x}_A \right\rangle \, ,
\end{align}
where $\mu_a$ are a collection of couplings that serve the same purpose as the chemical potentials in the analysis of the classical partition function in section \ref{sec:classical_partition_function}. 

For simplicity, we define $T = t_B - t_A$, and we will suppress the arguments of the propagator in what follows. One has the relations
\begin{align}
    \partial_\lambda K &= \left\langle \vec{x}_B \; \Big\vert \; - i T \widehat{\fO} \, e^{- i \widehat{H}_\lambda T + \mu_a \widehat{Q}_a } \; \Big\vert \; \vec{x}_A \right\rangle \, , \nonumber \\
    \partial_T K &= \left\langle \vec{x}_B \; \Big\vert \; - i \widehat{H} \, e^{- i \widehat{H}_\lambda T + \mu_a \widehat{Q}_a }\; \Big\vert \; \vec{x}_A \right\rangle \, , \nonumber \\
    \partial_{\mu_b} K &= \left\langle \vec{x}_B \; \Big\vert \; \widehat{Q}_b \, e^{- i \widehat{H}_\lambda T + \mu_a \widehat{Q}_a } \; \Big\vert \; \vec{x}_A \right\rangle \, .
\end{align}
Summation is implied in the expression $\mu_a \widehat{Q}_a = \sum_a \mu_a \widehat{Q}_a$. Because the deforming operator $\fO ( \widehat{H} , \widehat{Q}_a )$ is itself a function of the operators $\widehat{H}$ and $\widehat{Q}_a$, we arrive at a differential equation
\begin{align}
    \partial_\lambda K = - i T \fO \left[ i \partial_T , \partial_{\mu_a} \right] K \, .
\end{align}
This is identical to the flow equation (\ref{partition_function_flow}) for the classical partition function after identifying $\beta = i T$. As in section \ref{sec:classical_partition_function}, the expression $\fO \left[ i \partial_T , \partial_{\mu_a} \right]$ is defined by expanding the operator $\fO$ in a power series in $\widehat{H}$ and the $\widehat{Q}_a$, then replacing each instance of $\widehat{H}$ with $i \partial_T$ and each instance of $\widehat{Q}_a$ with $\partial_{\mu_a}$.

We may also think of this flow equation in terms of a generalization of the prescription of \cite{Gross:2019ach,Gross:2019uxi} for quantizing theories which are deformed by functions of the Hamiltonian. To do this, let us first recall how to derive the kernel representation of the propagator. Suppose that we can identify a complete basis of simultaneous eigenstates of the Hamiltonian $\widehat{H}_\lambda$ and each of the charge operators $\widehat{Q}_a$. We write these simultaneous eigenstates as $\ket{\phi_n}$, which satisfy
\begin{align}
    \widehat{H}_\lambda \ket{\phi_n} = E_n ( \lambda ) \ket{\phi}_n \, , \qquad \widehat{Q}_a \ket{\phi_n} = q_{a, n} \ket{\phi_n} \, .
\end{align}
In this case, we are thinking of the label $n$ as a multi-index which collects the quantum numbers for all of the charges. Using the completeness relation
\begin{align}
    \sum_n \ket{\phi_n} \bra{\phi_n} = 1 \, , 
\end{align}
we can evaluate the propagator as
\begin{align}\label{deformed_kernel_propagator}
    K &= \left\langle \vec{x}_B \; \Big\vert \;  e^{- i \widehat{H}_\lambda T + \mu_a \widehat{Q}_a } \; \Big\vert \; \vec{x}_A \right\rangle \nonumber \\
    &= \sum_{n, m} \langle \vec{x}_B \mid \phi_n \rangle \left\langle \phi_n \; \Big\vert \; e^{- i \widehat{H}_\lambda T + \mu_a \widehat{Q}_a } \; \Big\vert \; \phi_m \right\rangle \langle \phi_m \mid \vec{x}_A \rangle \nonumber \\
    &= \sum_{n, m} e^{- i E_n ( \lambda ) T + \mu_a q_{n, a} } \langle \vec{x}_B \mid \phi_n \rangle \delta_{m, n} \langle \phi_m \mid \vec{x}_A \rangle \nonumber \\
    &= \sum_n \phi_n^\ast ( \vec{x}_A ) \phi_n ( \vec{x}_B ) e^{- i E_n ( \lambda )  T + \mu_a q_{n, a} } \, .
\end{align}
Here we have used the orthogonality relation $\langle \phi_n \mid \phi_m \rangle = \delta_{m, n}$, along with the definition of the position space wavefunction
\begin{align}\label{position_space_wavefunction}
    \langle \vec{x} \mid \phi_n \rangle = \phi_n ( \vec{x} ) \, .
\end{align}
Equation (\ref{deformed_kernel_propagator}) suggests a straightforward interpretation of a deformation by conserved charges. In the deformed theory, every eigenstate $\ket{\phi_n}$ of the Hamiltonian and charges remains an eigenstate of the Hamiltonian and charges. The eigenvalues of $\ket{\phi_n}$ under each of the charge operators remains unchanged, since we deform the Hamiltonian $\widehat{H}$ but not the operators $\widehat{Q}_a$. It is easy to see that this is consistent with the fact that all of the operators $\widehat{Q}_a$ remain conserved in the deformed theory, since
\begin{align}
    \left[ \widehat{\fO} \, , \, \widehat{Q}_a \right] = 0 \, , 
\end{align}
by virtue of the fact that we assume all of the charges $\widehat{Q}_a$ are commuting (which is the quantum version of our assumption (\ref{poisson_commute}) that the charges are classically Poisson-commuting). All that has changed is that each ket $\ket{\phi_n}$ now has a deformed energy eigenvalue $E_n ( \lambda )$ which obeys the differential equation
\begin{align}\label{energy_eigenvalue_flow}
    \partial_\lambda E_n ( \lambda ) = \fO ( E_n, q_{n, a} ) \, , 
\end{align}
where on the right side of equation (\ref{energy_eigenvalue_flow}), $\fO$ is now a classical variable which is evaluated on the energy eigenvalue $E_n$ and charge eigenvalues $q_{n, a}$ of the state $\ket{\phi_n}$.

For instance, in the case of the the deformation by $\fO = \frac{H_\lambda^2}{\frac{1}{2} - 2 \lambda H_\lambda}$ whose classical solution was discussed around (\ref{1d_TT_flow_defn}), each energy eigenstate $\ket{\phi_n}$ with undeformed energy $E_n ( 0 ) $ remains an energy eigenstate in the deformed theory, but with a new energy
\begin{align}
    E_n ( \lambda ) = \frac{1}{4 \lambda} \left( 1 - \sqrt{ 1 - 8 \lambda E_n ( 0 )  } \right) \, .
\end{align}
There is a sharp difference in the behavior of the deformed spectrum depending on the sign of $\lambda$. If $\lambda < 0$, then the argument of the square root remains positive for arbitrary large positive undeformed energies. This choice is called the ``good sign'' of the deformation parameter. However, if $\lambda > 0$, then for sufficiently large undeformed energies $E_n ( 0 )$, the deformed energy levels become complex. This is the ``bad sign'' of the deformation. The same qualitative behavior occurs for the $\TT$ deformation of a $2d$ CFT.\footnote{In this setting, the complex spectrum can sometimes be returned to a purely real spectrum by performing sequential $\TT$ deformations with sufficiently large positive flow parameter \cite{Ferko:2022dpg}.}

\subsubsection*{\ul{\it Path Integral Analysis}}

We will now see how the above flow equations can be derived using the path integral. For the same reasons as we mentioned above, if the deforming operator $\fO$ depends on charges $Q_a$ besides the Hamiltonian, we cannot obtain a flow equation for the propagator using the unflavored path integral (\ref{phase_space_path_integral}). Instead we must introduce a flavored version which includes sources for the various charge operators,
\begin{align}\label{flavored_phase_space_path_integral}
    &K ( \vec{x}_B, t_B ; \vec{x}_A, t_A ; \lambda ; \mu_a ) \nonumber \\
    &\quad = \int_{\vec{x} ( t_A ) = \vec{x}_A}^{\vec{x} ( t_B ) = \vec{x}_B} \, \mathcal{D} \vec{x} \, \int_{\vec{p} ( t_A ) = \vec{p}_A}^{\vec{p} ( t_B ) = \vec{p}_B } \, \mathcal{D} \vec{p} \, \exp \left( i \int_{t_A}^{t_B} dt \, \left( p^i ( t ) \dot{x}_i ( t ) - H_\lambda ( \vec{p}, \vec{x} ) \right) + \sum_a \mu_a Q_a \right) \, .
\end{align}
To ease notation, we will again suppress the arguments of the propagator, omit the upper and lower bounds of path integration (which are always understood to take the values in (\ref{flavored_phase_space_path_integral})), and write $\mu_a Q_a$ for $\sum_a \mu_a Q_a$.

Let us emphasize two points about the phase space path integral expression for the propagator. First, as is typical of path integals, all dynamical quantities appearing inside the integrand are simply classical variables $x^i ( t )$ and $p^i ( t )$ rather than quantum operators $\hat{x}^i$ and $\hat{p}^i$, so there are no ordering ambiguities. Second, and more importantly, it is critical that the momentum variables $p^i ( t )$ inside the path integral are \emph{not} the conjugate momenta to $x^i ( t )$. The path integral runs over all phase space trajectories with the specified endpoints, and there is no constraint between the functions $x(t)$ and $p(t)$ along these trajectories. This is important because it implies that
\begin{align}
    \frac{\partial p^i ( t )}{\partial \lambda} = \frac{\partial \dot{x}^i ( t ) }{\partial \lambda} = 0 \, .
\end{align}
Let us contrast this situation with that of the proof of Theorem \ref{commute_theorem}. In that context, the momenta and velocities were related by equation (\ref{hamilton_eom}), the Hamilton equation of motion:
\begin{align}
    \frac{\partial H_\lambda}{\partial p^i} = \dot{x}^i \, .
\end{align}
Therefore, if we choose to treat $p^i$ as an independent variable which does not depend on $\lambda$, it follows that
\begin{align}\label{no_such_term}
    \frac{\partial \dot{x}^i}{\partial \lambda} = \frac{\partial ( \partial_\lambda H_\lambda )}{\partial p^i} \, , 
\end{align}
which is non-zero in general. This additional term appeared in (\ref{proof_reverse_intermediate}), where it was needed to demonstrate the equivalence of deformations of the Lagrangian and of the Hamiltonian.

However, since the $p^i$ and $x^i$ are unrelated integration variables in (\ref{flavored_phase_space_path_integral}), no term of the form (\ref{no_such_term}) is generated when we differentiate the propagator with respect to $\lambda$. The only $\lambda$ dependence appears in the Hamiltonian itself, so one finds
\begin{align}
    \partial_\lambda K &= \int \, \mathcal{D} \vec{x} \, \int \, \mathcal{D} \vec{p} \, \left( - i \int_{t_A}^{t_B} \, dt \, \fO \right) \exp \left( i \int_{t_A}^{t_B} dt \, \left( p^i ( t ) \dot{x}_i ( t ) - H_\lambda ( \vec{p} , \vec{x} ) \right) + \mu_a Q_a \right) \, .
\end{align}
We now use the fact that $\fO$ is only a function of conserved charges, which are independent of time. This means that the path integral expectation value of $\fO$ is itself also independent of time, so we may interchange the time integral with the path integral to conclude
\begin{align}
    \partial_\lambda K = - i ( t_B - t_A ) \langle \fO \rangle \, , 
\end{align}
where we have defined the path integral expectation value
\begin{align}
    \langle f ( x^j , p^j ) \rangle = \int \, \mathcal{D} \vec{x} \, \int \, \mathcal{D} \vec{p} \, f ( x^j , p^j ) \exp \left( i \int_{t_A}^{t_B} dt \, \left( p^i ( t ) \dot{x}_i ( t ) - H_\lambda ( \vec{p} , \vec{x} ) \right) + \mu_a Q_a  \right) \, .
\end{align}
Likewise, defining $T = t_B - t_A$, one has
\begin{align}\label{schrodinger}
    \partial_T K &= \left\langle i \left( p^i ( t ) \dot{x}_i ( t ) - H_\lambda ( p, x ) \right) \right\rangle \nonumber \\
    &= - i \langle H_\lambda \rangle \, .
\end{align}
Here we have used that the quantity $p(t) \dot{x}(t)$ is odd in both $p^i(t)$ and $x^i(t)$, and a path integral of an odd quantity over all paths vanishes by symmetry. Therefore the first term in the path integral expectation value of the first line of (\ref{schrodinger}) vanishes; the second term, $H_\lambda$, is constant in time because the Hamiltonian is conserved. Alternatively, one can justify the conclusion (\ref{schrodinger}) using the Schr\"odinger equation, which relates the time derivative of the propagator to the Hamiltonian. This Schr\"odinger relation for the phase space path integral is identical to that of the familiar Feynman path integral.

Similarly, the derivatives of the propagator with respect to the chemical potentials $\mu_a$ generate expectation values of charges:
\begin{align}\label{path_integral_mu_charge}
    \partial_{\mu_a} K = \langle Q_a \rangle \, .
\end{align}
Note that we have chosen conventions for the flavored phase space path integral (\ref{flavored_phase_space_path_integral}) such that the terms $\mu_a Q_a$ do not appear under the time integral. This leads to flow equations that are most similar to the classical results obtained for the grand canonical partition function (\ref{grand_canonical_defn}). However, because all of the charges $Q_a$ are conserved in time, we could have alternatively defined the path integral so that the terms $\mu_a Q_a$ were instead included inside of the time integral. This would have introduced an additional factor of $T$ in (\ref{path_integral_mu_charge}), and dependence on the charges in the computation of $\partial_T K$, which would produce a different flow equation that is related to our result by a redefinition of parameters.

We conclude that the expressions for $\partial_\lambda K$, $\partial_T K$, and $\partial_{\mu_a} K$ computed using the path integral formulation are identical to those computed using the canonical analysis. The phase space path integral representation for the propagator therefore obeys the same differential equation,
\begin{align}
    \partial_\lambda K = - i T \fO \left[ i \partial_T , \partial_{\mu_a} \right] K \, .
\end{align}
where the right side is defined for any analytic function $\fO$ of conserved charges.

\subsubsection*{\ul{\it Comments on Non-Analytic Deformations}}

Our derivation of the phase space path integral in section \ref{sec:phase_space_path_integral} assumed that the Hamiltonian operator $\widehat{H}$ admits a power series expansion in the operators $\hat{x}^i$ and $\hat{p}^i$. We also derived differential equations obeyed by the propagator, in either the canonical or path integral formalism, which involve expressions $\fO \left[ i \partial_T , \partial_{\mu_a} \right]$ that are defined by series expanding the deforming operator $\fO$ and replacing instances of $H$ and $Q_a$ with various derivatives. We now consider cases in which the deformation is not an analytic function of the charges, which includes the case of the $1d$ root-$\TT$ operator (\ref{rTT_1d}).

For a totally general non-analytic Hamiltonian $\widehat{H}$, it is not clear how to perform an analogue of the time-slicing prescription of section \ref{sec:phase_space_path_integral} and obtain a path integral definition. However, for a first-order deformation of an analytic Hamiltonian by a non-analytic function of conserved charges, the arguments of the preceding subsections apply in almost exactly the same way. At the risk of repeating ourselves, let us quickly check that this is true. We will consider a Hamiltonian with a form that is slightly more general than a first-order deformation, namely
\begin{align}\label{f1_and_f2_form}
    \widehat{H} = f_1 ( \lambda ) \widehat{H}_0 + f_2 ( \lambda ) \widehat{\fO} \, ,  
\end{align}
where $\widehat{H}_0$ is analytic and $\widehat{\fO}$ is a (possibly non-analytic) function of charges. When $f_1 = 1$ and $f_2 = \lambda$, this is the leading-order correction to $\widehat{H}_0$ generated by a flow driven by $\fO$.

As before, using the kernel representation of the propagator
\begin{align}\label{kernel_later}
    K = \sum_n \phi_n^\ast (x_A) \phi_n (x_B) e^{- i E_n T + \mu_a q_{n, a} } \, ,
\end{align}
one would similarly argue that a deformation of $\widehat{H}$ by a non-analytic function $\fO$ of conserved charges has the effect of leaving all of the eigenfunctions $\phi_n$ unchanged, and only shifts the energy eigenvalues as $E_n ( \lambda ) = f_1 ( \lambda ) E_n ( 0 ) + f_2 ( \lambda ) \fO$. One can still differentiate,
\begin{align}\label{non_analytic_dlambdaK}
    \partial_\lambda K = \sum_n \phi_n^\ast (x_A) \phi_n (x_B) \left( - i T \left( f_1' ( \lambda ) E_n ( 0 ) + f_2' ( \lambda ) \fO  \right) \right) e^{- i E_n ( \lambda ) T + \mu_a q_{n, a} } \, ,
\end{align}
although it may no longer be possible to express (\ref{non_analytic_dlambdaK}) in terms of derivatives of $K$ with respect to $T$ and the $\mu_a$ when $\fO$ is non-analytic.

A similar analysis is possible using the path integral. By the Baker–Campbell–Hausdorff formula, one has
\begin{align}\label{BCH_split}
    \exp \left( f_1 ( \lambda ) \widehat{H}_0 + f_2 ( \lambda ) \widehat{\fO} \right) = \exp \left( f_1 ( \lambda ) \widehat{H}_0 \right)  \exp \left( f_2 ( \lambda ) \widehat{\fO} \right) \, , 
\end{align}
since all commutator terms in the BCH expansion vanish by virtue of the fact that $\fO$ is a function only of conserved quantities. Suppose that we repeat the time-slicing prescription of section \ref{sec:phase_space_path_integral} for this Hamiltonian. When evaluating (\ref{time_sliced_propagator}), one has
\begin{align}
    \langle \vec{x}_{j+1} \mid \widehat{U} ( t_{j+1}, t_{j} ) \mid \vec{x}_{j} \rangle &= \langle \vec{x}_{j+1} \mid e^{- i f_1 ( \lambda ) \widehat{H}_0 \epsilon} e^{- i f_2 ( \lambda ) \widehat{\fO} \epsilon} \mid \vec{x}_{j} \rangle \nonumber \\
    &= \sum_{n, m} \langle \vec{x}_{j+1} \mid e^{- i f_1 ( \lambda ) \widehat{H}_0 \epsilon} \mid \phi_n \rangle \langle \phi_n \mid e^{- i f_2 ( \lambda ) \widehat{\fO} \epsilon} \mid \phi_m \rangle \langle \phi_m \mid \vec{x}_{j} \rangle \, , 
\end{align}
where we have inserted two complete sets of eigenstates $\ket{\phi_n}$ of both the undeformed Hamiltonian $\widehat{H}_0$ and the charge operators $\widehat{Q}_a$. We can then evaluate the middle factor by replacing $\widehat{\fO}$ with its classical value $\fO$, giving
\begin{align}
    \langle \vec{x}_{j+1} \mid \widehat{U} ( t_{j+1}, t_{j} ) \mid \vec{x}_{j} \rangle &= \sum_{n, m} e^{- i f_2 ( \lambda ) \fO \epsilon} \langle \vec{x}_{j+1} \mid e^{- i f_1 ( \lambda ) \widehat{H}_0 \epsilon} \mid \phi_n \rangle \langle \phi_n  \mid \phi_m \rangle \langle \phi_m \mid \vec{x}_{j} \rangle \nonumber \\
    &= e^{- i f_2 ( \lambda ) \fO \epsilon} \langle \vec{x}_{j+1} \mid e^{- i f_1 ( \lambda ) \widehat{H}_0 \epsilon} \mid \vec{x}_{j} \rangle \, , 
\end{align}
and then we may evaluate the remaining matrix element for the analytic part $f_1 ( \lambda ) \widehat{H}_0$ of the Hamiltonian using the same steps as before. This would lead us to the same result,
\begin{align}\label{propagator_first_order}
    \partial_\lambda K = \left\langle - i T \left( f_1' ( \lambda ) \widehat{H}_0 + f_2' ( \lambda ) \widehat{\fO} \right) \right\rangle \, .
\end{align}
Taking $f_1 = 1$ and $f_2 = \lambda$ allows us to conclude, using either formalism, that a deformation of an analytic Hamiltonian by a non-analytic function of charges -- to leading order in the deformation parameter -- is described by the same differential equation $\partial_\lambda K = \langle - i T \widehat{\fO} \rangle$ for the propagator. We emphasize that the only difference in the case of a non-analytic deforming operator is that the expectation value $\langle \widehat{\fO} \rangle$ may not be expressible in terms of derivatives of the propagator with respect to $T$ and the $\mu_a$.

One might then ask how one can extend this analysis to higher orders in the deformation parameter. For instance, suppose that we use the analysis above to define the Hamiltonian $\widehat{H}^{(1)} = \widehat{H}_0 + \lambda \widehat{\fO}$ which solves the flow equation $\partial_\lambda \widehat{H}_\lambda = \widehat{\fO}$ to first order, and we then wish to treat $\widehat{H}^{(1)}$ as a new seed Hamiltonian to deform again and generate the second-order solution $\widehat{H}^{(2)}$, and continue in this way to define all higher $\widehat{H}^{(n)}$. Let us make two comments on this point.

\begin{enumerate}[label=(\Roman*)]
    \item The first comment is that a general, first-principles analysis of this process would require a procedure for obtaining a path integral representation for the propagator of an arbitrary non-analytic seed Hamiltonian, which is not available. Instead, we can simply \emph{define} what we mean by the all-orders version of the deformed quantum theory by first solving the differential equation $\partial_\lambda H_\lambda = \fO$, and then inserting this solution for $H_\lambda$ into the phase space path integral (\ref{final_phase_space_path_integral}).
    
    By construction, this is equivalent to using the corresponding solution for the deformed energy levels $E_n ( \lambda )$ and inserting them into the kernel representation (\ref{kernel_later}) of the propagator. In this way, we obtain a consistent prescription for the quantization of a non-analytic theory which agrees with the above analysis to leading order around an analytic seed Hamiltonian.

    \item\label{non_analytic_okay} The second comment is that, in some cases of interest, we can sidestep the issue of performing path integral quantization of a non-analytic seed Hamiltonian. Specifically, for the deformation of the harmonic oscillator by the $1d$ root-$\TT$ operator to obtain the ModMax oscillator, it turns out that the \emph{all-orders} solution to the flow equation takes the form (\ref{f1_and_f2_form}) for appropriately chosen functions $f_1$ and $f_2$. In this case, the analysis which we carried out for the leading-order deformation is sufficient to derive differential equations that hold to all orders along the flow.
\end{enumerate}

\subsection{Flow of Quantum Partition Function and Comparison to $2d$} \label{sec:compare_flows_2d}

In the preceding subsection, we have developed general differential equations obeyed by the propagator $K ( \vec{x}_B, t_B ; \vec{x}_A, t_A )$ of a quantum theory deformed by a function of conserved charges. Since the Euclidean time propagator with periodic boundary conditions is the thermal partition function, which can also be written in the trace form
\begin{align}
    Z ( \beta ) = \Tr \left( e^{- \beta \widehat{H} } \right) \, , 
\end{align}
one can likewise obtain differential equations obeyed by $Z ( \beta )$, or more precisely for the grand canonical partition function with chemical potentials for the various charges:
\begin{align}\label{quantum_grand_canonical}
    \mathcal{Z} ( \beta, \mu_1 , \ldots , \mu_M ) = \Tr \left( e^{- \beta \widehat{H} + \sum \mu_a \widehat{Q}_a } \right) \, .
\end{align}
One can derive flow equations for (\ref{quantum_grand_canonical}) using manipulations which are identical to those around equation (\ref{partition_function_flow}). In particular,
\begin{align}
    \partial_\beta \mathcal{Z} = \Tr \left( - \widehat{H}  e^{- \beta \widehat{H} + \sum \mu_a \widehat{Q}_a } \right) = - \langle \widehat{H} \rangle \, , \qquad \partial_{\mu_a} \mathcal{Z} = \Tr \left( \widehat{Q}_a e^{- \beta \widehat{H} + \sum \mu_b \widehat{Q}_b } \right) = \langle \widehat{Q}_a \rangle \, , 
\end{align}
and thus
\begin{align}\label{quantum_partition_function_flow}
    \partial_\lambda \mathcal{Z} = - \beta \fO \left[ - \partial_\beta , \partial_{\mu_1} , \ldots , \partial_{\mu_M} \right] \left(  \mathcal{Z} \right) \, .
\end{align}
In fact, we could have derived the flow equation for the classical partition function by first arguing that the quantum partition function satisfies the differential equation (\ref{quantum_partition_function_flow}) and then taking the limit $\hbar \to 0$.

We will discuss some examples of such flow equations only briefly, because they take the same form as the corresponding flow equations for the classical partition function described in section \ref{sec:classical_partition_function}. However, since these differential equations now hold in the quantum theory, we will comment on the relationship to the analogous flow equations for the quantum mechanical partition functions of two-dimensional field theories which are deformed by the $\TT$ operator.

For instance, under the flow (\ref{1d_TT_example}) which is the $1d$ version of the $\TT$ deformation, the propagator $K$ and thermal partition function $Z$ obey the differential equations
\begin{align}\label{1d_quantum_Z_TT}
    \left( 4 \lambda \partial_\lambda \partial_T + 2 T \partial_T^2 + i \left( 1 + \frac{4 i \lambda}{T} \right) \partial_\lambda \right) K_\lambda ( T ) &= 0 \, , \nonumber \\
    \left( 4 \lambda \partial_\lambda \partial_\beta + 2 \beta \partial_\beta^2 + \left( 1 - \frac{4 \lambda}{\beta} \right) \partial_\lambda \right) Z_\lambda ( \beta ) &= 0 \, .
\end{align}
This flow equation for $Z$ was also considered in \cite{Iliesiu:2020zld}. It can be understood as follows. Suppose that we begin with a two-dimensional conformal field theory whose torus partition function is $Z_0 ( \tau , \overbar{\tau} )$, where $\tau$ is the modular parameter of the torus. One can then deform this theory by the $\TT$ operator to obtain a one-parameter family of deformed torus partition functions $Z_\lambda$ which obey the differential equation
\begin{align}\label{2d_TT_partition_function_flow}
    \partial_\lambda Z_\lambda ( \tau, \overbar{\tau} ) = \left( \tau_2 \partial_{\tau} \partial_{\overbar{\tau}} + \frac{1}{2} \left( \partial_{\tau_2} - \frac{1}{\tau_2} \right) \lambda \partial_\lambda \right) Z_\lambda ( \tau, \overbar{\tau} ) \, .
\end{align}
This differential equation can be obtained by performing a Hubbard-Stratonovich transformation which replaces the $\TT$ deformation with a random metric \cite{Cardy:2018sdv}. Modular properties of the deformed partition function were studied in \cite{Datta:2018thy,Aharony:2018bad}; although the $\TT$-deformed theory no longer enjoys conformal symmetry, the partition function $Z_\lambda$ is still invariant with respect to a modular transformation under which the parameter $\lambda$ also transforms.

Let us specialize to a torus with purely imaginary modular parameter $\tau = \frac{i \beta}{8}$. Using
\begin{align}
    \partial_{\tau} = \frac{1}{2} \left( \partial_{\tau_1} - i \partial_{\tau_2} \right) \, , \qquad \partial_{\overbar{\tau}} = \frac{1}{2} \left( \partial_{\tau_1} + i \partial_{\tau_2} \right) \, , 
\end{align}
with $\tau_2 = \frac{\beta}{8}$ and $\tau_1 = 0$, the flow equation (\ref{2d_TT_partition_function_flow}) reduces to
\begin{align}
    \partial_\lambda Z_\lambda ( \beta ) = \left( 2 \beta \partial_\beta^2 + 4 \left( \partial_\beta - \frac{1}{\beta} \right) \lambda \partial_\lambda \right) Z_\lambda ( \beta ) \, , 
\end{align}
which reproduces (\ref{1d_quantum_Z_TT}) after sending $\lambda \to - \lambda$, which is a choice of conventions for the deformation parameter. Thus the $1d$ $\TT$-deformed partition function indeed descends from the $2d$ $\TT$-deformed partition function.

It is interesting to consider a similar dimensional reduction of the two-dimensional root-$\TT$ flow. The analogous flow equation for the torus partition function of a root-$\TT$ deformed $2d$ CFT was conjectured in \cite{Ebert:2023tih} to be
\begin{align}\label{Z_flow_root_TT}
    \partial_\gamma^2 Z_\gamma ( \tau , \overbar{\tau } )  =  \left( \tau_2^2 \partial_\tau \partial_{\overbar{\tau}} + \tau_2 \partial_{\tau_2} \right) Z_\gamma ( \tau , \overbar{\tau } ) \, ,
\end{align}
based on a proposal for the flow of the cylinder spectrum which was justified by holographic considerations. If we take a torus with modular parameter
\begin{align}
    \tau = i \beta + \mu \, , 
\end{align}
then the flow equation (\ref{Z_flow_root_TT}) reduces to 
\begin{align}\label{root_TT_thermal_partition_function_flow}
    \left( \partial_\gamma^2 - \beta \partial_\beta - \frac{\beta^2}{4} \left( \partial_\beta^2 + \partial_\mu^2  \right) \right) \mathcal{Z} ( \beta, \gamma, \mu )  = 0 \, .
\end{align}
The factor of $\frac{1}{4}$ multiplying the third term in (\ref{root_TT_thermal_partition_function_flow}) is related to the normalization of the root-$\TT$ operator, and can be rescaled to $1$ by an appropriate redefinition. Up to this choice of scaling, this is the same differential equation which is satisfied by the partition function of the ModMax oscillator, which we will present in equation (\ref{almost_cylinder}).

\section{Application to ModMax Oscillator}\label{sec:modmax_osc}

In this section, we will apply the general results of the preceding sections to the main example of interest in the present work, which is the ModMax oscillator. This theory was first introduced in \cite{Garcia:2022wad} and is a particular deformation of an isotropic harmonic oscillator. Given a collection of position variables $x^i$, for $i = 1 , \ldots, N$, we begin by defining the undeformed theory with the Lagrangian
\begin{align}
    L_0 = \frac{1}{2} \left( \dot{x}^i \dot{x}^i - x^i x^i \right) \, , 
\end{align}
where we have set the mass $m$ and frequency $\omega$ of the harmonic oscillator to $1$ for convenience. This theory has a conserved energy
\begin{align}
    E_0 = \frac{\partial L_0}{\partial \dot{x}^i} \dot{x}^i - L_0 = \frac{1}{2} \left( \dot{x}^i \dot{x}^i + x^i x^i \right) \, , 
\end{align}
which is the Noether current associated with time translation symmetry, along with a collection of conserved angular momenta
\begin{align}
    J^{nm}_0 = \frac{\partial L_0}{\partial \dot{x}^n} x^m - \frac{\partial L_0}{\partial \dot{x}^m} x^n = \dot{x}^n x^m - \dot{x}^m x^n \, , 
\end{align}
which are the conserved currents associated with rotations $x^i \to \tensor{R}{^i_j} x^j$, $R \in SO(N)$. The total angular momentum is
\begin{align}
    J_0^2 = J^{nm}_0 J_{0, nm} \, .
\end{align}
For any $\gamma$, we can now define the Lagrangian for the ModMax oscillator as
\begin{align}\label{modmax_osc_lagrangian}
    L_\gamma = \frac{\cosh ( \gamma )}{2} \left( \dot{x}^i \dot{x}^i - x^i x^i \right) \pm \sinh ( \gamma ) \sqrt{ E_0^2 - J_0^2 } \, .
\end{align}
Note that there is a choice of the relative sign between the two terms in (\ref{modmax_osc_lagrangian}) which is correlated with the choice of sign for the root-$\TT$ operator that drives the flow equation (\ref{1d_rTT_flow_lagrangian}) below. One can also view this sign choice as a convention for the sign of the parameter $\gamma$, since sending $\gamma \to - \gamma$ reverses the relative sign.

Even though this Lagrangian $L_\gamma$ is written in terms of the conserved quantities $E_0$ and $J_0^2$ in the \emph{undeformed} theory, it satisfies a differential equation which involves the conserved currents in the \emph{deformed} theory at finite $\gamma$, namely
\begin{align}
    E_\gamma = \frac{\partial L_\gamma}{\partial \dot{x}^i} \dot{x}^i - L_\gamma \, , \qquad J^{nm}_\gamma = \frac{\partial L_\gamma}{\partial \dot{x}^n} x^m - \frac{\partial L_\gamma}{\partial \dot{x}^m} x^n \, , \qquad J_\gamma^2 = J^{nm}_\gamma J_{\gamma, nm} \, .
\end{align}
One can show that $L_\gamma$ obeys
\begin{align}\label{1d_rTT_flow_lagrangian}
    \frac{\partial L_\gamma}{\partial \gamma} = \pm \sqrt{ E_\gamma^2 - J_\gamma^2 } \, ,
\end{align}
which we refer to as the $1d$ root-$\TT$ flow equation. The corresponding Hamiltonian,
\begin{align}\label{modmax_osc_hamiltonian}
    H_\gamma = \frac{\cosh ( \gamma )}{2} \left( p^i p^i  + x^i x^i \right) \mp \sinh ( \gamma ) \sqrt{ E_0^2 - J_0^2 } \, ,
\end{align}
satisfies a flow equation with the opposite sign,
\begin{align}\label{1d_rTT_flow_hamiltonian}
    \frac{\partial H_\gamma}{\partial \gamma} = \mp \sqrt{ E_\gamma^2 - J_\gamma^2 } \, ,
\end{align}
as required by Theorem \ref{commute_theorem}. In equations (\ref{modmax_osc_hamiltonian}) and (\ref{1d_rTT_flow_hamiltonian}), the quantities $E_0$, $E_\gamma$, $J_0^2$, and $J_\gamma^2$ are defined by beginning with the appropriate Noether currents computed in the Lagrangian formulation, and then expressing these quantities in terms of conjugate momenta $p^i$ rather than velocities $\dot{x}^i$.

Although the ModMax oscillator can be defined for any number $N$ of position variables $x^i$, $i = 1 , \ldots, N$, we will focus on the case $N = 2$ for simplicity. First let us review some features of the classical dynamics of the ModMax oscillator.

\subsection{Classical Aspects}

We now specialize to the case of $N = 2$ coordinates $x^i$, and we will use the notation $x^1 = x$, $x^2 = y$. In terms of these variables, the general Lagrangian (\ref{modmax_osc_lagrangian}) for the ModMax oscillator can be written as
\begin{align}\label{2d_modmax_oscillator_lagrangian} \hspace{-7pt}
    L_\gamma = \frac{1}{2} \left[ \cosh ( \gamma ) \left( \dot{x}^2 + \dot{y}^2 - x^2 - y^2 \right) \pm \sinh ( \gamma ) \sqrt{ \left( \left( y + \dot{x} \right)^2 + \left( x - \dot{y} \right)^2 \right) \left( \left( x + \dot{y} \right)^2 + \left( y - \dot{x} \right)^2 \right) } \right] \, .
\end{align}
The conserved angular momentum, which is the Noether charge associated with rotations in the $(x, y)$ plane, is
\begin{align}
    J_\gamma = \left( x \dot{y} - y \dot{x} \right) \cosh ( \gamma ) \pm \sinh ( \gamma ) \cdot \frac{\left( x \dot{y} - y \dot{x} \right) \left( \dot{x}^2 + \dot{y}^2 - x^2 - y^2 \right) }{\sqrt{ \left( \left( y + \dot{x} \right)^2 + \left( x - \dot{y} \right)^2 \right) \left( \left( x + \dot{y} \right)^2 + \left( y - \dot{x} \right)^2 \right) }} \, .
\end{align}
It is interesting to express $J_\gamma$ in terms of the conjugate momenta $p_x$ and $p_y$, as appropriate for formulating flows for the Hamiltonian. The conjugate momenta computed from the Lagrangian (\ref{2d_modmax_oscillator_lagrangian}) are
\begin{align}
    p_x &= \frac{\partial L_\gamma}{\partial \dot{x}} = \dot{x} \cosh ( \gamma ) \pm \sinh ( \gamma ) \cdot \frac{2 x y \dot{y} + x^2 \dot{x} + \dot{x} \left( \dot{x}^2 + \dot{y}^2 - y^2 \right) }{\sqrt{ \left( \left( y + \dot{x} \right)^2 + \left( x - \dot{y} \right)^2 \right) \left( \left( x + \dot{y} \right)^2 + \left( y - \dot{x} \right)^2 \right) }} \, , \nonumber \\
    p_y &= \frac{\partial L_\gamma}{\partial \dot{y}} = \dot{y} \cosh ( \gamma ) \pm \sinh ( \gamma ) \cdot \frac{ 2 x y \dot{x} - x^2 \dot{y} + \dot{y} \left( \dot{x}^2 + \dot{y}^2 + y^2 \right) }{\sqrt{ \left( \left( y + \dot{x} \right)^2 + \left( x - \dot{y} \right)^2 \right) \left( \left( x + \dot{y} \right)^2 + \left( y - \dot{x} \right)^2 \right) }} \, .
\end{align}
After expressing the angular momentum $J_\gamma$ in terms of $p_x$ and $p_y$, one finds
\begin{align}\label{deformed_J_undeformed_J}
    J_\gamma = x p_y - y p_x \, .
\end{align}
That is, when written in Hamiltonian variables, the deformed angular momentum $J_\gamma$ takes the same functional form as the undeformed angular momentum $J_0$. This is a special case of the observation, which we first made in the text below equation (\ref{position_space_wavefunction}), that $f ( H , Q_a )$ deformations modify the Hamiltonian but not the other charges such as $J$.

Similarly, for $N=2$ one can write the Hamiltonian for the ModMax oscillator as
\begin{align}\label{2d_modmax_hamiltonian}\hspace{-10pt}
    H_\gamma = \frac{1}{2} \left[ \cosh ( \gamma ) \left( p_x^2 + p_y^2 + x^2 + y^2 \right) \mp \sinh ( \gamma ) \sqrt{ \left( \left( p_y + x \right)^2 + \left( p_x - y \right)^2 \right) \left( \left( p_y - x \right)^2 + \left( p_x + y \right)^2 \right) } \right] \, .
\end{align}
Let us consider the symmetries of this theory in somewhat more detail. It is well-known that the undeformed theory, which is the $2d$ isotropic harmonic oscillator, enjoys an $SU(2)$ symmetry. To see this, it is convenient to define complex variables
\begin{align}
    z = x + i p_x \, , \qquad w = y + i p_y \, , 
\end{align}
so that the harmonic oscillator Hamiltonian is
\begin{align}
    H_0 = \frac{1}{2} \left( | z |^2 + | w |^2 \right) \, .
\end{align}
This Hamiltonian is invariant under any action of the form
\begin{align}
    \begin{bmatrix} z \\ w \end{bmatrix} \to U \begin{bmatrix} z \\ w \end{bmatrix} \, , \qquad U \in SU ( 2 ) \, , 
\end{align}
since such an $SU(2)$ transformation preserves the length of the complex vector. In these complex variables, the angular momentum is
\begin{align}
    J = \frac{1}{2i} \left( w \overbar{z} - z \overbar{w} \right) = \Im ( w \overbar{z} ) \, .
\end{align}
The angular momentum $J$ is \emph{not} invariant under the full $SU(2)$ symmetry group. However, it is still invariant under the restricted $U(1)$ transformations
\begin{align}\label{complex_phase}
    z \to e^{i \alpha} z \, , \qquad w \to e^{i \alpha} w \, , \qquad \alpha \in \mathbb{R} \, .
\end{align}
Similarly, any deformed Hamiltonian which is a function of both the undeformed Hamiltonian and this angular momentum,
\begin{align}
    H = H ( H_0, J ) \, , 
\end{align}
is also invariant under the $U(1)$ transformations (\ref{complex_phase}).

We have commented before that the ModMax oscillator is a particular dimensional reduction of the four-dimensional ModMax theory, which enjoys electric-magnetic duality invariance. In fact, the $U(1)$ invariance (\ref{complex_phase}) of the ModMax oscillator descends directly from this electric-magnetic duality symmetry, which can be written as
\begin{align}
    z_{\mu \nu} \to e^{i \alpha} z_{\mu \nu} \, , \qquad z_{\mu \nu} = F_{\mu \nu} + i \widetilde{F}_{\mu \nu} \, ,
\end{align}
where $F_{\mu \nu}$ is the field strength of the $4d$ electrodynamics theory and $\widetilde{F}_{\mu \nu}$ is its Hodge dual.

It is straightforward to see that any deformation of the isotropic harmonic oscillator which is constructed from the Hamiltonian and the conserved angular momentum will preserve invariance under the $U(1)$ duality transformation (\ref{complex_phase}). This is the $1d$ version of the statement that any deformation of a theory of self-dual electrodynamics in four spacetime dimensions, where the deforming operator is a function of the energy-momentum tensor of the theory, will preserve electric-magnetic duality invariance.  See \cite{Ferko:2023wyi} for futher discussion and examples of such duality-preserving stress tensor flows.

\subsubsection*{\ul{\it Flow Equation for Partition Functions}}

As we have seen in sections \ref{sec:classical} and \ref{sec:quantum}, both the classical and quantum partition functions for a theory deformed by a function of conserved charges satisfy the same differential equation. We will now study this differential equation in the case of the ModMax oscillator, which obeys a flow driven by the operator $\mathcal{R}$ or $\mathfrak{R}$ introduced in equation (\ref{rTT_1d}). This falls into the class of non-analytic deformations which we briefly considered around equation (\ref{square_root_Blow}). In this case, the differential equation (\ref{second_order_Z_flow_sqrt}) simplifies considerably.

The reason for this simplification is the following. We have seen that the solution to the flow equation in this case is given by the Lagrangian (\ref{2d_modmax_oscillator_lagrangian}) or Hamiltonian (\ref{2d_modmax_hamiltonian}), which satisfy the equations
\begin{align}
    \frac{\partial^2 L_\gamma}{\partial \gamma^2} = L_\gamma \, , \qquad \frac{\partial^2 H_\gamma}{\partial \gamma^2} = H_\gamma \, .
\end{align}
Because $\partial_\gamma L_\gamma = \pm \mathcal{R}^{(\gamma)}$ and $\partial_\gamma H_\gamma = \mp \mathfrak{R}^{(\gamma)}$, this means
\begin{align}
    \frac{\partial \mathcal{R}^{(\gamma)}}{\partial \gamma} = \pm L_\gamma \, , \qquad \frac{\partial \mathfrak{R}^{(\gamma)}}{\partial \gamma} = \mp H_\gamma \, .
\end{align}
The flow equation (\ref{second_order_Z_flow_sqrt}) then becomes
\begin{align}\label{almost_cylinder}
    \partial_\gamma^2 \mathcal{Z} &= \left\langle - \beta H_\gamma + \beta^2 \left( H_\gamma^2 - J_\gamma^2 \right) \right\rangle \nonumber \\
    &= \beta \partial_\beta \mathcal{Z} + \beta^2 \partial_\beta^2 \mathcal{Z} - \beta^2 \partial_\mu^2 \mathcal{Z} \, ,
\end{align}
where in the last step we have expressed quantities in terms of derivatives of the partition function. Note that, if we had instead chosen a different normalization for the $1d$ root-$\TT$ operator, so that the flow equation for the Lagrangian were $\partial_\gamma L_\gamma = c_0 \mathcal{R}^{(\gamma)}$ for some constant $c_0$, the flow equation would have been
\begin{align}
    \partial_\gamma^2 \mathcal{Z} - \beta \partial_\beta \mathcal{Z} - \beta^2 c_0^2 \left( \partial_\beta^2 \mathcal{Z} - \partial_\mu^2 \mathcal{Z} \right) = 0 \, .
\end{align}
For the choice $c_0 = \frac{1}{2}$, this matches the dimensional reduction of the conjectured flow equation for the torus partition function of a root-$\TT$ deformed CFT given in  (\ref{root_TT_thermal_partition_function_flow}).

Equation (\ref{almost_cylinder}) is very nearly of a familiar form. To see this, it is convenient to Wick-rotate $\mu \to i \mu$ and $\gamma \to i \gamma$, which reverses the signs on two terms. The resulting differential equation can be written as
\begin{align}\label{wick_rotated_Z_flow}
    0 = \frac{1}{\beta} \partial_\beta \left( \beta \partial_\beta \mathcal{Z} \right) + \frac{1}{\beta^2} \partial_\gamma^2 \mathcal{Z} + \partial_\mu^2 \mathcal{Z} \, .
\end{align}
This is identical to the Laplace equation for a function $f : \mathbb{R}^3 \to \mathbb{R}$ written in cylindrical coordinates $(r, \theta, z)$, namely
\begin{align}
    0 &= \frac{1}{r} \frac{\partial}{\partial r} \left( r \frac{\partial f}{\partial r} \right) + \frac{1}{r^2} \frac{\partial^2 f}{\partial \theta^2} + \frac{\partial^2 f}{\partial z^2} \, ,
\end{align}
where the roles of the coordinates $(r, \theta, z)$ are played by $(\beta, \gamma, \mu)$, respectively.

One may therefore solve the flow equation for $\mathcal{Z} ( \beta, \gamma, \mu )$ by separation of variables, in a manner analogous to that which is done when studying electrodynamics in cylindrical coordinates. The original differential equation (\ref{almost_cylinder}), before Wick-rotating the parameters $\mu$ and $\gamma$, has different signs than those which appear in the cylindrical Laplace equation, leading to the appearance of slightly different Bessel functions. Any function of the form
\begin{align}
    f_{a, b} ( \beta, \gamma, \mu ) = \left( c_1 e^{- a \gamma} + c_2 e^{a \gamma} \right) \left( c_3 e^{- b \mu} + c_4 e^{b \mu} \right) \left( c_5 I_a ( b \beta ) + c_6 K_a ( b \beta ) \right) \, , 
\end{align}
for constants $a$, $b$, and $c_1$, $\ldots$, $c_6$, is a solution to (\ref{almost_cylinder}). Here $I_\nu ( x )$ and $K_\nu ( x )$ are modified Bessel functions of the first and second kind, respectively. A general solution to the flow equation will therefore be a sum or integral of such functions $f_{a, b}$ for various choices of $a$ and $b$. Physical considerations will also restrict the choices of the parameters $c_i$. For instance, the modified Bessel functions of the second kind $K_\nu ( x )$ are generically divergent as $x \to 0$, whereas we expect a partition function to be finite at $\beta = 0$, so one should normally set $c_6 = 0$ in cases of physical interest.

We note that formal analytic continuation of flow equations to imaginary values of the parameters, like that which relates (\ref{almost_cylinder}) and (\ref{wick_rotated_Z_flow}), has sometimes been useful in previous work. For instance, in \cite{Ferko:2023ruw} such a continuation of a $\TT$-like parameter $\lambda$ was useful in relating the flow equations which produce the $4d$ Born-Infeld and reverse-Born-Infeld theories, which are two of the four solutions to the zero-birefringence condition for $4d$ nonlinear electrodynamics \cite{Russo:2022qvz}.

\subsubsection*{\ul{\it Direct Computation of Classical Partition Function}}

As a warm-up for our study of the quantum partition function, and in order to illustrate an example of the Laplace equation which the deformed flavored partition functions satisfy, we will now perform a direct computation of the classical partition function for the $2d$ ModMax oscillator. The resulting formulas will turn out to be tidier if we evaluate this partition function with an imaginary chemical potential for the angular momentum, which merely reverses a sign in the corresponding flow equation. We will therefore compute
\begin{align}
    \mathcal{Z} ( \beta, \gamma , \mu )  = \frac{1}{(2 \pi)^2 } \int d x \, d y \, d p_x \, d p_y \, \exp \left( - \beta H_\gamma + i \mu J_\gamma \right) \, ,
\end{align}
where all integrals run from $- \infty$ to $\infty$. Writing this integrand explicitly in terms of the positions and momenta, and making the sign choice for which the two terms in the Hamiltonian $H_\gamma$ are both manifestly positive, the integral we wish to evaluate is
\begin{align}\label{Z_classical}
    \mathcal{Z} ( \beta, \gamma , \mu ) &= \frac{1}{(2 \pi)^2 } \int d x \, d y \, d p_x \, d p_y \, \exp \Bigg[ - \beta \Bigg( \frac{1}{2} \cosh ( \gamma ) \left( p_x^2 + p_y^2 + x^2 + y^2 \right) \nonumber \\
    &+ \frac{1}{2} \sinh ( \gamma ) \sqrt{ \left( \left( p_y + x \right)^2 + \left( p_x - y \right)^2 \right) \left( \left( p_y - x \right)^2 + \left( p_x + y \right)^2 \right) } \Bigg ) + i \mu \left( x p_y - y p_x \right) \Bigg] \, .
\end{align}
Note that we have used the expression $J_\gamma = x p_y - y p_x$ in equation (\ref{Z_classical}) since the deformed angular momentum in Hamiltonian variables takes the same form as the undeformed angular momentum, as we pointed out around equation (\ref{deformed_J_undeformed_J}).

It is convenient to perform the change of variables
\begin{align}
    u_1 = p_y + x \, , \quad u_2 = p_x - y \, , \quad v_1 = p_y - x \, , \quad v_2 = p_x + y \, ,
\end{align}
so that the integral becomes
\begin{align}
    \mathcal{Z} ( \beta, \gamma , \mu ) &= \frac{1}{4 (2 \pi)^2 } \int d u_1 \, d u_2 \, d v_1 \, d v_2 \, \exp \Bigg[ - \frac{\beta}{4} \Bigg( \left( u_1^2 + u_2^2 + v_1^2 + v_2^2 \right) \nonumber \\
    &\quad + 2 \sinh ( \gamma ) \sqrt{ u_1^2 + u_2^2 } \sqrt{ v_1^2 + v_2^2 } \Bigg ) + \frac{i \mu}{4} \left( u_1^2 + u_2^2 - v_1^2 - v_2^2 \right) \Bigg] \, .
\end{align}
Note that this change of variables has decoupled the square root interaction into two factors. We can now go to polar coordinates in the $(u_1, u_2)$ and $(v_1, v_2)$ planes as
\begin{align}
    u_1 = r_u \cos ( \theta_u ) \, , \quad u_2 = r_u \sin ( \theta_u ) \, , \quad v_1 = r_v \cos ( \theta_v ) \, , \quad v_2 = r_v \sin ( \theta_v ) \, .
\end{align}
Then our partition function is
\begin{align}
    \mathcal{Z} ( \beta, \gamma , \mu ) &= \frac{1}{4 (2 \pi)^2 } \int_{0}^{\infty} \, d r_u \int_{0}^{\infty} \, d r_v \int_{0}^{2 \pi} \, d \theta_u \int_{0}^{2 \pi} \, d \theta_v \, \, r_u \, r_v \, \exp \Bigg[ - \frac{\beta}{4} \Bigg( \cosh ( \gamma ) \left( r_u^2 + r_v^2 \right) \nonumber \\
    &\quad + 2 \sinh ( \gamma ) r_u r_v \Bigg ) + \frac{i \mu}{4} \left( r_u^2 - r_v^2 \right) \Bigg] \, .
\end{align}
The angular integrals give factors of $2 \pi$, whereas the resulting radial integrals can be evaluated in closed form, and we find
\begin{align}\label{classical_partition_gamma_mu}
    \mathcal{Z} ( \beta, \gamma , \mu ) = \frac{1}{ \beta^2 + \mu^2  } \left( 1 - \frac{\sinh ( \gamma ) }{\sqrt{ 1 + \frac{\mu^2}{\beta^2} } } \arctan \left( \csch ( \gamma ) \sqrt{ 1 + \frac{\mu^2}{\beta^2} } \right) \right) \, .
\end{align}
This is our final expression for the classical grand canonical partition function for the $2d$ ModMax oscillator at inverse temperature $\beta$ and with imaginary chemical potential $i \mu$ for the angular momentum. One can of course obtain the result for the opposite sign choice in the Hamiltonian by sending $\gamma \to - \gamma$. The result (\ref{classical_partition_gamma_mu}) clearly reduces to the corresponding partition function for the ordinary $2d$ harmonic oscillator when $\gamma = 0$. One can also check by explicit computation that it obeys the partial differential equation
\begin{align}
    \left( \partial_\gamma^2 - \beta^2 \partial_\beta^2 - \beta \partial_\beta - \beta^2 \partial_\mu^2 \right) \mathcal{Z} ( \beta, \gamma , \mu ) = 0 \, ,
\end{align}
which is equivalent to (\ref{almost_cylinder}) after Wick-rotating the chemical potential $\mu \to i \mu$. One can recover the corresponding solution with real chemical potential by reversing the signs of all instances of $\mu^2$ in equation (\ref{classical_partition_gamma_mu}).

This result gives one example of a partition function which satisfies a flow equation driven by a non-analytic combination of charges, namely the $1d$ root-$\TT$ deformation. When the chemical potential is set to zero, the deformed partition function (\ref{classical_partition_gamma_mu}) is simply a $\gamma$-dependent rescaling of the undeformed partition function. However, when $\mu$ and $\gamma$ are both finite, the temperature dependence is modified in a more interesting way. We will see shortly that, in the quantum theory, even for $\mu = 0$ the deformed partition function is not simply a rescaling of the undeformed partition function.

\subsection{Quantum Aspects}

We now turn to the main subject of this work, which is the quantum mechanics of the ModMax oscillator. At first glance, it is not so clear that one should be able to quantize this theory at all. Na\"ively, one would like to begin with the classical Hamiltonian (\ref{2d_modmax_hamiltonian}) and promote all position and momentum variables to operators. This requires one to make sense of the operator square root in the second term, which takes the form
\begin{align}\label{naive_square_root_operator}
    \sqrt{ \left( \left( \hat{p}_y + \hat{x} \right)^2 + \left( \hat{p}_x - \hat{y} \right)^2 \right) \left( \left( \hat{p}_y - \hat{x} \right)^2 + \left( \hat{p}_x + \hat{y} \right)^2 \right) } \, .
\end{align}
It is not immediately obvious what this operator should mean. First note that we would not expect that it is possible to define an operator square root for a generic combination of position and momentum operators, at least without additional assumptions like positivity. For instance, the expression $\sqrt{ \hat{x} }$ does not give a conventional Hermitian operator; even if one attempts to define it by diagonalizing the position operator and declaring $\sqrt{ \hat{x} } \ket{x} = \sqrt{ x } \ket{x}$, this operator will have imaginary eigenvalues for negative positions $x$. 

In our case, we are aided in interpreting the operator (\ref{naive_square_root_operator}) by the fact that it is positive definite -- which allows us to define it by diagonalization and taking square roots -- and because it is a function of conserved charges in the undeformed theory, which allows us to write flow equations for quantities in the deformed theory using the results of section \ref{sec:quantum}. We will begin by attempting to understand the operator (\ref{naive_square_root_operator}) directly using raising and lowering operators in the theory of the undeformed harmonic oscillator.

\subsubsection*{\ul{\it Ladder Operator Representation}}

We can develop one useful perspective on the ModMax oscillator by rewriting the Hamiltonian in terms of creation and annihilation operators. As usual, when studying the undeformed theory of an isotropic $2d$ harmonic oscillator with Hamiltonian
\begin{align}
    \widehat{H}_0 = \frac{1}{2} \left( \hat{p}_x^2 + \hat{p}_y^2 + \hat{x}^2 + \hat{y}^2 \right) \, , 
\end{align}
it is natural to define the annihilation operators
\begin{align}
    \hat{a}_x = \frac{1}{\sqrt{2}} \left( \hat{x} + i \hat{p}_x \right) \, , \qquad \hat{a}_y = \frac{1}{\sqrt{2}} \left( \hat{y} + i \hat{p}_y \right) \, , 
\end{align}
whose Hermitian conjugates are the creation operators,
\begin{align}
    \hat{a}_x^\dagger = \frac{1}{\sqrt{2}} \left( \hat{x} - i \hat{p}_x \right) \, , \qquad \hat{a}_y^\dagger = \frac{1}{\sqrt{2}} \left( \hat{y} - i \hat{p}_y \right) \, .
\end{align}
In terms of these operators the Hamiltonian takes the standard form
\begin{align}
    \widehat{H}_0 = 1 + \hat{a}_x^\dagger \hat{a}_x + \hat{a}_y^\dagger \hat{a}_y = 1 + \widehat{N}_x + \widehat{N}_y \, , 
\end{align}
where we have defined the number operators $\widehat{N}_i = \hat{a}_i^\dagger \hat{a}_i$. However, the angular momentum operator can be made more transparent by a change of basis. We can instead define the ``circularly-polarized'' linear combinations
\begin{align}\label{left_right_ladders}
    \hat{a}_L = \frac{1}{\sqrt{2}} \left( \hat{a}_x + i \hat{a}_y \right) \, , \qquad \hat{a}_R = \frac{1}{\sqrt{2}} \left( \hat{a}_x - i \hat{a}_y \right) \, .
\end{align}
The corresponding number operators, $\widehat{N}_{L} = \hat{a}_{L}^\dagger \hat{a}_{L}$ and $\widehat{N}_{R} = \hat{a}_{R}^\dagger \hat{a}_{R}$, count the numbers of left-moving and right-moving circular quanta. This is a useful way to leverage the rotational symmetry of the problem, since the angular momentum operator is now simply
\begin{align}
    \widehat{J} = \widehat{N}_R - \widehat{N}_L \, , 
\end{align}
and the Hamiltonian is
\begin{align}
    \widehat{H}_0 = 1 + \widehat{N}_L + \widehat{N}_R \, .
\end{align}
We now turn to the $1d$ root-$\TT$ deformation which generates the interaction term in the ModMax oscillator Lagrangian. This operator is
\begin{align}
    \widehat{\mathfrak{R}} &= \sqrt{ \widehat{H}_0^2 - \widehat{J}^2 } \nonumber \\
    &= \sqrt{ \left( 1 + 2 \widehat{N}_L \right) \left( 1 + 2 \widehat{N}_R \right) } \, .
\end{align}
The argument of the square root factorizes into left-moving and right-moving pieces. Because the left-moving and right-moving operators commute, we are free to split the square root into separate factors:
\begin{align}
    \widehat{\mathfrak{R}} = \sqrt{ 1 + 2 \widehat{N}_L } \sqrt{ 1 + 2 \widehat{N}_R} \, .
\end{align}
Each of the operators $1 + 2 \widehat{N}_L$ and $1 + 2 \widehat{N}_R$ have strictly positive eigenvalues, and it is therefore possible to define an operator square root. This is equivalent to defining the square root operators by the Taylor series expansions
\begin{align}\label{root_taylor}
    \sqrt{1 + 2 \widehat{N}_L} = \sum_{k = 0}^{\infty} \binom{\frac{1}{2}}{k} \left( 2 \widehat{N}_L \right)^k \, , \qquad \sqrt{1 + 2 \widehat{N}_R} = \sum_{k = 0}^{\infty} \binom{\frac{1}{2}}{k} \left( 2 \widehat{N}_R \right)^k \, .
\end{align}
Either of these infinite sums is convergent and well-defined when acting on any state in the Hilbert space. We can therefore write the Hamiltonian of the ModMax oscillator as
\begin{align}\label{H_gamma_numbers}
    \widehat{H}_\gamma = \cosh ( \gamma ) \left( 1 + \widehat{N}_L + \widehat{N}_R \right) + \sinh ( \gamma ) \sqrt{ 1 + 2 \widehat{N}_L } \sqrt{ 1 + 2 \widehat{N}_R } \, ,
\end{align}
where we have again chosen the positive sign for concreteness, although one can obtain the other sign choice by taking $\gamma$ to be negative.

One could study the quantum mechanics of this theory in essentially the same way that we have described above for the canonical quantization prescription. That is, one considers a complete basis of eigenstates $\ket{ N_L, N_R }$ of the undeformed Hamiltonian and angular momentum, and then shifts each of their energy eigenvalues according to (\ref{H_gamma_numbers}).

However, this number operator representation also allows us to see a simple way to generate the ModMax oscillator in one step from the undeformed isotropic harmonic oscillator. Let us introduce operators
\begin{align}
    \widehat{M}_L = \sqrt{ 1 + 2 \widehat{N}_L } \, , \qquad \widehat{M}_R = \sqrt{ 1 + 2 \widehat{N}_R } \, , 
\end{align}
which are defined by the convergent Taylor series expansions (\ref{root_taylor}). Then the undeformed Hamiltonian can be written as
\begin{align}
    \widehat{H}_0 = \frac{1}{2} \left( \widehat{M}_L^2 + \widehat{M}_R^2 \right) \, .
\end{align}
Suppose that one now performs the transformation
\begin{align}\label{transformation}
    \widehat{M}_L \to \widehat{M}_L^{(\gamma)} = \cosh ( \gamma ) \widehat{M}_L + \sinh ( \gamma ) \widehat{M}_R \, , \qquad \widehat{M}_R \to \widehat{M}_R^{(\gamma)} = \cosh ( \gamma ) \widehat{M}_R + \sinh ( \gamma ) \widehat{M}_L \, .
\end{align}
Then one finds
\begin{align}
    \widehat{H}_0 \to \widehat{H}_{2 \gamma} &=\frac{1}{2} \left( \left( \widehat{M}_L^{(\gamma)} \right)^2 + \left( \widehat{M}_R^{(\gamma)} \right)^2 \right) \, , \nonumber \\
    &= \cosh ( 2 \gamma ) \left( 1 + \widehat{N}_R + \widehat{N}_L \right) + \sinh ( 2 \gamma ) \sqrt{ 1 + 2 \widehat{N}_L } \sqrt{ 1 + 2 \widehat{N}_R } \, .
\end{align}
This is exactly the deformed Hamiltonian of equation (\ref{H_gamma_numbers}), except at parameter $2 \gamma$ rather than $\gamma$. Therefore, performing a ``boost'' in the space of the operators $\widehat{M}_L$ and $\widehat{M}_R$ has the effect of generating the ModMax oscillator Hamiltonian in a single step, rather than via a flow equation which is defined by infinitesimally deforming by the operator $\widehat{\mathfrak{R}}$.

This is the same structure as the classical deformation map which was introduced in section 6 of \cite{Garcia:2022wad}. In that case, a particular redefinition of positions and momenta had the effect of mapping $H_0 \to H_{2 \gamma}$ in a single step, albeit at the level of the classical Hamiltonian. The transformation (\ref{transformation}) on the number-like operators $\widehat{M}_L$ and $\widehat{M}_R$ can be thought of as a quantum version of this deformation map.

\subsubsection*{\ul{\it Trace Form of Quantum Partition Function}}

We now consider the grand canonical partition function for the ModMax theory at the quantum level. Following the ladder operator representation of the Hamiltonian developed above, it is convenient to choose a basis of simultaneous eigenstates $\ket{N_L, N_R}$ of the number operators associated with the left- and right-circularly-polarized creation and annihilation operators introduced in equation (\ref{left_right_ladders}).

Using this basis, the flavored partition function with an imaginary chemical potential for the angular momentum $\widehat{J} = \widehat{N}_R - \widehat{N}_L$ can be written as
\begin{align}\label{quantum_Z}
    \mathcal{Z} ( \beta, \gamma, \mu ) &= \Tr \left( \exp \left( - \beta \widehat{H}_\gamma + i \mu \widehat{J}_\gamma \right) \right) \nonumber \\
    &= \sum_{N_L, N_R = 0}^{\infty} \exp \Big[ - \beta \left( \cosh ( \gamma ) \left( 1 + \widehat{N}_R + \widehat{N}_L \right) + \sinh ( \gamma ) \sqrt{ 1 + 2 \widehat{N}_R } \sqrt{ 1 + 2 \widehat{N}_L } \right) \nonumber \\
    &\qquad \qquad \qquad \qquad + i \mu \left( \widehat{N}_R - \widehat{N}_L \right) \Big] \, .
\end{align}
Here we have used the fact that $\widehat{J}_\gamma = \widehat{J}_0 = \widehat{N}_R - \widehat{N}_L$, following the comments around (\ref{deformed_J_undeformed_J}), and taken the positive sign choice in the Hamiltonian as usual.

Unlike in the case of the classical partition function, it does not seem to be possible to obtain a simple closed-form expression for the sum (\ref{quantum_Z}). However, it is straightforward to evaluate the trace perturbatively in the flow parameter $\gamma$. For instance, to leading order one finds
\begin{align}\label{Z_quantum_leading}
    \mathcal{Z} ( \beta, \gamma, \mu ) = \frac{1}{2 \cosh ( \beta ) - 2 \cos ( \mu ) } - 2 \gamma \beta e^{\beta} \Phi \left( e^{-\beta - i \mu} , - \frac{1}{2} , \frac{1}{2} \right) \Phi \left( e^{- \beta + i \mu} , - \frac{1}{2} , \frac{1}{2} \right) + \mathcal{O} \left( \gamma^2 \right) \, , 
\end{align}
where $\Phi ( z, s, a)$ is a special function known as the Lerch transcendent and defined by
\begin{align}
    \Phi ( z, s, a ) = \sum_{k=0}^{\infty} \frac{z^k}{ ( k + a )^s } \, .
\end{align}
Even when $\mu = 0$, the expression (\ref{Z_quantum_leading}) for the partition function to order $\gamma$ is not simply a rescaling of the undeformed partition function by a $\gamma$-dependent factor. This is unlike the classical partition function (\ref{classical_partition_gamma_mu}), to which the expression (\ref{quantum_Z}) reduces in the limit $\hbar \to 0$.\footnote{This fact is not obvious from equation (\ref{quantum_Z}) because we have set $\hbar = 1$ for simplicity. After restoring factors of $\hbar$ and taking $\hbar \to 0$, the sum reduces to the integral (\ref{Z_classical}), as it must.} This gives one way to see that the quantum theory of the ModMax oscillator is richer than its classical counterpart, since even without a chemical potential $\mu$ there is a non-trivial interplay between the inverse temperature $\beta$ and flow parameter $\gamma$.

One can check that the infinite sum (\ref{quantum_Z}) satisfies the flow equation (\ref{almost_cylinder}) for a theory deformed by the $1d$ root-$\TT$ operator, after specializing to an imaginary value of $\mu$. As we mentioned, this differential equation can solved by separation of variables, which gives a general solution that is a sum of factorized terms involving exponentials and Bessel functions. It is instructive to see how (\ref{quantum_Z}) can be brought into this form, since as written this sum is not obviously related to Bessel functions. This can be accomplished using a variant of the Jacobi-Anger expansion, which expresses a plane wave as a superposition of cylindrical waves. For any $z, \theta \in \mathbb{C}$, these identities take the form
\begin{align}\label{jacobi_anger}
    e^{z \cos ( \theta ) } &= I_0 ( z ) + 2 \sum_{k=1}^{\infty} I_k ( z ) \cos ( k \theta ) \, \nonumber \\
    e^{z \sin ( \theta ) } &= I_0 ( z ) + 2 \sum_{k=0}^{\infty} ( - 1 )^k I_{2 k + 1} ( z ) \sin \left( \left( 2 k + 1 \right) \theta \right) + 2 \sum_{k=1}^{\infty} ( - 1 )^k I_{2k} ( z ) \cos ( 2 k \theta ) \, .
\end{align}
See, for instance, section 10.35 of \cite{olver10}. To apply these identities to the partition function (\ref{quantum_Z}) for the quantum ModMax oscillator, we let $\theta = i \gamma$ in the identities (\ref{jacobi_anger}). The full partition function can then be written as the expansion
\begin{align}
    &\mathcal{Z} ( \beta, \gamma, \mu ) \nonumber \\
    &= \sum_{N_L, N_R = 0}^{\infty} \Bigg\{ \Big[ \cos \left( \mu ( N_R - N_L ) \right) + i \sin \left( \mu ( N_R - N_L ) \right) \Big] \nonumber \\
    &\qquad \cdot \left[ I_0 \left( - \beta ( 1 + N_R + N_L ) \right) + 2 \sum_{k=1}^{\infty} I_k \left( - \beta ( 1 + N_R + N_L ) \right) \cosh ( k \gamma ) \right] \nonumber \\
    &\qquad \cdot \Bigg[ I_0 \left( - i \beta \sqrt{ 1 + 2 N_R} \sqrt{ 1 + 2 N_L } \right) + 2 \sum_{k=1}^{\infty} ( - 1 )^k I_{2k} \left( - i \beta \sqrt{ 1 + 2 N_R} \sqrt{ 1 + 2 N_L } \right)  \cosh ( 2 k \gamma )  \nonumber \\
    &\qquad \qquad + 2 i \sum_{k=0}^{\infty} ( - 1 )^k I_{2 k + 1} \left( - i \beta \sqrt{ 1 + 2 N_R} \sqrt{ 1 + 2 N_L } \right)  \sinh \left( \left( 2 k + 1 \right) \gamma \right) \Bigg] \Bigg\} \, .
\end{align}
Although rather unwieldy, this expansion in Bessel functions makes the connection between the trace form (\ref{quantum_Z}) of the partition function, and the Laplace-type equation which it satisfies, more explicit.

\subsubsection*{\ul{\it Description of Deformed Propagator}}

To conclude, we will comment on the characterization of the propagator for the ModMax oscillator. Here we will be brief, since this is a direct application of the general results of section \ref{sec:quantum} and leads to flow equations which are essentially identical to those for the classical and quantum partition function discussed above.

Because the Hamiltonian for the ModMax oscillator is of the form (\ref{f1_and_f2_form}), there is no ambiguity in defining the propagator by a path integral representation, even at all orders in $\gamma$. Therefore, the flavored propagator is defined by the general phase space path integral given in equation (\ref{flavored_phase_space_path_integral}). This propagator, with real chemical potential, satisfies
\begin{align}\label{propagator_pde}
    \partial_\gamma^2 K - T \partial_T K - T^2 \left( \partial_T^2 + \partial_\mu^2 \right) K = 0 \, ,
\end{align}
which is the same as the differential equation for the flavored partition function after making the replacement $\beta = i T$. However, we emphasize that the result is more general, since this holds for the propagator $K ( \vec{x}_B, t_B ; \vec{x}_A, t_A ; \lambda ; \mu )$ with any initial position $\vec{x}_A$ and final position $\vec{x}_B$. In contrast, the thermal partition function is obtained from the Euclidean time propagator with periodic boundary conditions, which means that the initial and final positions are equal.

The differential equation (\ref{propagator_pde}) fully determines the propagator for the ModMax oscillator, given the initial condition $K ( \gamma = 0 )$ and the first derivative $\partial_\gamma K \vert_{\gamma = 0}$, which is related to the expectation value of the $1d$ root-$\TT$ operator in the undeformed theory. The initial condition $K ( \gamma = 0 )$ is essentially the propagator for a $2d$ harmonic oscillator in a background magnetic field, which plays the role of the chemical potential for the angular momentum. This quantity can be computed either by canonical methods or path integral methods; for the path integral computation, see the pedagogical review \cite{10.1119/1.17809}.

We also note that the propagator can be written using the kernel representation and the basis of states $\ket{N_L, N_R}$, which gives
\begin{align}
    K ( \vec{x}_B, t_B ; \vec{x}_A, t_A ; \lambda ; \mu ) &= \sum_{N_L, N_R = 0}^{\infty} \phi_{N_L, N_R}^\ast ( \vec{x}_A ) \phi_{N_L, N_R} ( \vec{x}_B ) \exp \Bigg[ - i \Bigg( \cosh ( \gamma ) \left( 1 + \widehat{N}_R + \widehat{N}_L \right) \nonumber \\
    &\qquad + \sinh ( \gamma ) \sqrt{ 1 + 2 \widehat{N}_R } \sqrt{ 1 + 2 \widehat{N}_L } \Bigg) ( t_B - t_A )  + \mu \left( \widehat{N}_R - \widehat{N}_L \right) \Bigg] \, , 
\end{align}
where the $\phi_{N_L, N_R}$ are harmonic oscillator wavefunctions, which are known in closed form.

Our characterization of the propagator $K$, including that it satisfies the Laplace-like differential equation (\ref{propagator_pde}), is one of the main results of this work. Because the propagator for the ModMax oscillator is completely determined by the above considerations, this essentially constitutes a full solution of the model. Any physical question involving time evolution of states can in principle be extracted from the function $K$. This therefore completes our study of the quantum mechanical theory of the ModMax oscillator.

\section{Conclusion}\label{sec:conclusion}

In this work, we have studied general deformations of $1d$ theories by conserved charges, both at the classical and quantum level. This has allowed us to obtain flow equations for quantities in the theory of the ModMax oscillator, which is the dimensional reduction of the $4d$ ModMax theory. In particular, we have found that the thermal partition function in the quantum theory of the ModMax oscillator -- or, relatedly, the real-time propagator -- satisfies a certain partial differential equation which is related by Wick-rotation to the Laplace equation in $3d$ cylindrical coordinates.

One way of summarizing our analysis is to say that \emph{any} deformation of a quantum mechanical theory by conserved charges is essentially ``solvable'' in the sense that one can write differential equations which relate quantities in the deformed theory, such as the propagator or partition function, to those in the undeformed theory. The results on the ModMax oscillator are a special case of this fact when the deformation is driven by the $1d$ root-$\TT$ operator. Furthermore, the quantization of such charge-deformed models is unambiguous, since one obtains equivalent flow equations using either canonical quantization or path integral quantization. We therefore conclude that the fairly broad class of theories obtained through deformations by conserved charges should be included among other examples of solvable deformations of quantum mechanics, such as the one in which the quadratic kinetic term is replaced by one involving a hyperbolic cosine \cite{Grassi:2018bci}.

There are several directions for future investigation, some of which we outline below.

\subsubsection*{\ul{\it One-Loop Calculation}}

In this manuscript, we have focused on finding \emph{exact} flow equations for observables in the quantum theory of the ModMax oscillator, such as the propagator. We have also studied certain quantities in the classical theory, such as the classical partition function, for which it is possible to obtain a closed-form expression.

However, it would be also interesting to study \emph{semi-classical} expressions for quantum observables by performing an expansion in $\hbar$. In the limit as $\gamma \to 0$, the ModMax oscillator reduces to the ordinary harmonic oscillator, which is one-loop exact. It seems very unlikely that the deformed theory is also one-loop exact due to the complicated nature of the interaction term. One could attempt to compute the one-loop correction around the classical solution to the equations of motion for the ModMax oscillator and examine how closely this reproduces the full quantum results.

To do this, one would need to expand the phase space path integral which defines the propagator for the ModMax oscillator and retain terms up to quadratic order in fluctuations around the classical path. Fortunately, it is straightforward to write down the general classical solution to the equations of motion for the ModMax oscillator following \cite{Garcia:2022wad}. Given a set of initial conditions $(x_0, y_0)$ and $(p_{x,0} , p_{y,0} )$, one can evaluate the conserved energy $H_0$ and angular momentum $J_0$ corresponding to this initial condition, 
\begin{align}
    E_0 = \frac{1}{2} \left( x^2 + y^2 + p_x^2 + p_y^2 \right) \, , \qquad J = x p_y - y p_x \, ,
\end{align}
and then define
\begin{align}
    A = \cosh ( \gamma ) - \frac{\sinh ( \gamma ) H_0}{\sqrt {H_0^2 - J_0^2}} \, , \qquad B = \frac{\sinh ( \gamma ) J_0}{\sqrt{ H_0^2 - J_0^2 } } \, .
\end{align}
The general solution to the deformed equations of motion is given by
\begin{align}\label{modmax_eom_soln}
    x ( t ) &= \sin ( A t ) \left( p_{x, 0} \cos ( B t ) + p_{y, 0} \sin ( B t ) \right) + \cos ( A t ) \left( x_0 \cos ( B t ) + y_0 \sin ( B t ) \right) \, , \nonumber \\
    y ( t ) &= \sin ( A t ) \left( p_{y, 0} \cos ( B t ) - p_{x, 0} \sin ( B t ) \right) + \cos ( A t ) \left( y_0 \cos ( B t ) - x_0 \sin ( B t ) \right) \, , \nonumber \\
    p_x ( t ) &= \cos ( A t ) \left( p_{x, 0} \cos ( B t ) + p_{y, 0} \sin ( B t ) \right) - \sin ( A t ) \left( x_0 \cos ( B t ) + y_0 \sin ( B t ) \right) \, , \nonumber \\
    p_y ( t ) &= \cos ( A t ) \left( p_{y, 0} \cos ( B t ) - p_{x, 0} \sin ( B t ) \right) - \sin ( A t ) \left( y_0 \cos ( B t ) - x_0 \sin ( B t ) \right) \, .
\end{align}
One could then perform a one-loop computation by defining
\begin{align}
    x^i = x^i_{\text{cl}} + \delta x^i \, , \qquad p^i = p^i_{\text{cl}} + \delta p^i \, , 
\end{align}
where $x^i_{\text{cl}} ( t )$ and $p^i_{\text{cl}} ( t )$ are a solution to the equations of motion (\ref{modmax_eom_soln}), and then performing the phase space path integral over $\delta p^i$ and $\delta x^i$. This is slightly more involved than a semiclassical computation in the ordinary Feynman path integral, which gives a one-loop determinant. In this case, one would first expand the Hamiltonian action to write
\begin{align}
    H = f_1 ( \delta x ) \left( \delta p^2 \right) + f_2 ( \delta x ) \delta p + f_3 ( \delta x ) \, ,
\end{align}
where we suppress indices on the fluctuations. It is still possible to carry out the $\mathcal{D} ( \delta p )$ path integral over momentum fluctuations for such a Hamiltonian, as described in section 1.4 of \cite{Mosel:2004mk}. After performing the momentum path integral, one is left with a path integral over $\mathcal{D} ( \delta x )$ with a modified Lagrangian. It should then be possible to complete the semi-classical expansion by computing the one-loop determinant associated with the fluctuations $\delta x$ around the classical solution in this Feynman path integral.

\subsubsection*{\ul{\it Laplace Equation for Flavored Partition Function and Narain Moduli Space}}

One of the main results of this manuscript is that the propagator for a root-$\TT$ deformed theory satisfies a partial differential equation which -- up to signs which can be eliminated by choosing imaginary values of parameters -- is identical to the Laplace equation in three-dimensional cylindrical coordinates. Similar Laplace-type equations have appeared in the description of certain torus partition functions for two-dimensional conformal field theories. For instance, in \cite{Datta:2021ftn} the authors study flavored CFT partition functions which satisfy a Laplace equation where the Laplacian acts both on a space of Narain lattices and on a space of chemical potentials. This is of the same schematic form as the Laplace equation which we have derived for the ModMax oscillator partition function.

Another observation along similar lines is the following. We have pointed out that the flow equation we obtained for a $1d$ root-$\TT$ deformed theory descends by dimensional reduction from the flow equation (\ref{Z_flow_root_TT}) for a root-$\TT$ deformed torus partition function. Equation (\ref{Z_flow_root_TT}) is structurally similar to the equation obeyed by certain theta functions. For instance, consider a theory of $D$ compact bosons which parameterize a target-space torus $T^D$ that has some collection of moduli which we schematically indicate by $m$. The partition function for this theory is
\begin{align}
    Z ( m , \tau ) = \frac{\Theta ( m , \tau ) }{ | \eta ( \tau ) |^{2D} } \, , 
\end{align}
where $\Theta ( m , \tau )$ is a Siegel-Narain theta function, which obeys the differential equation
\begin{align}\label{narain_PDE}
    \left( - \tau_2^2 \partial_\tau \partial_{\overbar{\tau}} - \tau_2 \partial_{\tau_2} - \Delta_{\mathcal{M}_D} \right) \Theta ( m , \tau ) = 0 \, ,
\end{align}
where $\Delta_{\mathcal{M}_D}$ is the natural Laplacian on the Narain moduli space that parameterizes the $T^D$. For instance, in the case of a single compact boson, the target space is a circle of radius $R$ and the Laplacian is
\begin{align}
    \Delta_{\mathcal{M}_1} = - \frac{1}{4} \left( R \frac{d}{dR} \right)^2 \, .
\end{align}
The structure of equation (\ref{Z_flow_root_TT}) is almost identical to (\ref{narain_PDE}), except with the Laplacian on moduli space replaced with the second derivative with respect to $\gamma$. Also note that the root-$\TT$ flow equation involves the partition function itself, while (\ref{narain_PDE}) holds for the function $\Theta$ appearing in the numerator of the partition function, not for the full combination including the eta function in the denominator.

It would be interesting to understand whether there is a deeper relationship between these flow equations for root-$\TT$ deformed partition functions, and their dimensional reductions to $1d$, and properties of Narain moduli space.

\subsubsection*{\ul{\it Coupling to Worldline Gravity}}

As we have emphasized, there may be several inequivalent quantization schemes -- or ``UV completions'' -- for a particular classical theory. In particular, this is true for theories obtained from the $(0+1)$-dimensional version of the $\TT$ deformation. One way to see the difference between two such choices of quantization scheme is by examining the resulting thermal partition functions. As we reviewed in section \ref{sec:classical_partition_function}, the quantization procedure for the $1d$ $\TT$ deformation which gives rise to the flow equation (\ref{TT_QM_partition_function_flow}) admits a solution for the deformed partition function which can be written as an integral transform of the undeformed partition function, equation (\ref{1d_TT_Z_integral_kernel}).

However, there is a second UV completion of this deformation which is defined by coupling the seed theory to worldline gravity \cite{Gross:2019ach}. In this prescription, the deformed partition function is defined by the path integral
\begin{align}\label{TT_worldline_gravity_path_integral}
    Z_\lambda ( \beta ) = \int \frac{\mathcal{D} e \, \mathcal{D} X \, \mathcal{D} \sigma}{\mathrm{Vol}} \exp \left( - S_0 ( e ; X ) - S ( \lambda ; e, \sigma ) \right) \, , 
\end{align}
where $X$ represents the fields of the undeformed theory, which are now minimally coupled to an einbein $e$, and $\sigma$ is an auxiliary scalar field with an action
\begin{align}
    S ( \lambda ; e , \sigma ) = \frac{1}{2 \lambda} \int_0^{\beta'} \, d \tau \, e \left( e^{-1} \partial_\tau \sigma - 1 \right)^2 \, .
\end{align}
After evaluating the path integral, one finds that this quantization scheme produces the deformed partition function
\begin{align}\label{worldline_gravity_TT_Z}
    Z_\lambda ( \beta ) = \int_0^{\infty} d \beta' \, \frac{ \beta }{\sqrt{2 \pi \lambda} \beta^{\prime 3/2}} \sum_{m} \exp \left( - \frac{ \left( m \beta - \beta' \right)^2 }{2 \beta' \lambda} \right) \, Z_0 ( \beta' ) \, ,
\end{align}
which is a different partition function than the result (\ref{1d_TT_Z_integral_kernel}), although they agree in the unit winding sector ($m=1$) up to normalization for the flow parameter.\footnote{In particular, one can redefine $\lambda \to - 4 \lambda$ in (\ref{1d_TT_Z_integral_kernel}) to match the conventions of (\ref{worldline_gravity_TT_Z}).}

It would be very interesting to find an analogue of this worldline gravity prescription for the $1d$ root-$\TT$-like deformation. It seems likely that one would need to couple the undeformed theory to both an einbein $e ( \tau )$ and a gauge field for the $SO(N)$ symmetry, which plays the role of a time-dependent magnetic field. This would be in accord with the fact that, in order to obtain a flow equation for the partition function using our simpler quantization prescription for the root-$\TT$ deformed quantum mechanics, we were forced to turn on a chemical potential $\mu$ for the angular momentum. Finding a path integral expression which represents this alternative, worldline gravity quantization prescription would allow us to better understand the available choices of UV completions for ModMax-like theories of quantum mechanics.

\section*{Acknowledgements}

We thank Stephen Ebert, Mukund Rangamani, Savdeep Sethi, Dmitri Sorokin, Zhengdi Sun, and Gabriele Tartaglino-Mazzucchelli for helpful discussions, and we thank Julian Malory for proofreading a draft of this manuscript. C. F. is supported by U.S. Department of Energy grant DE-SC0009999 and by funds from the University of California.

\appendix

\section{First-Order Analysis of Lagrangian and Hamiltonian Flows}\label{app:first_order}

In this appendix, we will consider the analogue of Theorem \ref{commute_theorem} where one uses the relationship between the conjugate momenta $p^i$ and velocities $\dot{x}^i$ in the \emph{undeformed} theory (defined by $L_0$ and $H_0$) rather than in the deformed theory (defined by $L_\lambda$ and $H_\lambda$). If one does not correct the definition of the conjugate momenta, the resulting deformations of the Lagrangian and Hamiltonian will be equivalent only to leading order in the deformation parameter. The logic of this proof was originally presented in appendix A of \cite{Kruthoff:2020hsi}, which focused on $\TT$ deformations of two-dimensional quantum field theories describing a field $\phi$ and conjugate momentum $\pi$. Here we will instead consider $1d$ theories which describe the dynamics of a collection of positions $x^i$ and velocities $\dot{x}^i$, since this is the primary focus of the present work, although the reasoning is almost identical.

Let us represent the deformation of the Hamiltonian to first order as follows:
\begin{align}
    H(x^{i}, p^{i})=H_{0}(x^{i}, p^{i}) + \epsilon \fO (x^{i}, p^{i}) \, .
\end{align}
We claim that this is equivalent to a deformation of the Lagrangian to first order as
\begin{align}\label{first_order_lagrangian_def}
    L(x^{i}, \Dot{x}^{i})=L_{0}(x^{i}, \Dot{x}^{i}) - \epsilon \fO ( x^i, f_0^i ( \dot{x}^j ) )  \, .
\end{align}
Here we define $f_{0}^i$ to be the function that relates $\dot{x}^{i}$ and $p^{i}$ when $\epsilon=0$. More precisely,
\begin{align}
    p^i = \frac{\partial L_0}{\partial \dot{x}^i} = f^i_0 ( \dot{x}^j ) \, .
\end{align}
Because we are defining the deformation in terms of the Hamiltonian, we view the momenta $p^i$ as independent variables which do not depend on the deformation parameter $\lambda$; this corresponds to the forward direction of Theorem \ref{commute_theorem}, rather than the converse. Consequently the definition of the velocities $\dot{x}^i$ changes due to the deformation.

We use the Legendre transform to write the deformed Lagrangian in terms of the Hamiltonian,
\begin{align}
    L(x^{i}, p^{i}) = p^{i}\Dot{x}^{i}-H_{0}(x^{i}, p^{i}) - \epsilon \fO ( x^i, p^i )  \, .
\end{align}
Next, we solve the equations of motion to write $p^{i}$ as a function of $x^{i}$ and $\Dot{x}^{i}$. We assume that this solution can be written as an infinite series as follows:
\begin{align}
    p^{i} = f^i ( x^{j}, \dot{x}^{j} ) = f_{0}^i ( x^{j}, \dot{x}^{j} ) + \epsilon f_{1}^i ( x^{j}, \Dot{x}^{j} ) + \mathcal{O} ( \epsilon^{2} ).
\end{align}
By Hamilton's equations of motion, $\Dot{x}^{i}=\frac{\partial H}{\partial p^{i}}$. We can use the equation for $p^{i}$ to first order to obtain,
\begin{align}
    \dot{x}^{i} = \frac{\partial H_{0}}{\partial p^{i}} ( x^{j}, f_{0}^j + \epsilon f_{1}^j ) + \epsilon \frac{\partial \fO}{\partial p^{i}} (x^{j}, f_{0}^j + \epsilon f_{1}^j ).
\end{align}
We expand the first term perturbatively to obtain the following expression for $\dot{x}^{i}$:
\begin{align}\label{appendix_compare_order_epsilon}
    \dot{x}^{i} = \frac{\partial H}{\partial p^{i}} ( x^{j}, f_{0}^j )  + \epsilon f_{1}^j \frac{\partial^{2}H_{0}}{\partial p^i \partial p^j} ( x^{k}, f_{0}^k ) + \epsilon \frac{\partial \fO}{\partial p^{i}} (x^{j}, f_{0}^j ) + \mathcal{O} ( \epsilon^{2} ).
\end{align}
Matching the terms of order $\epsilon^0$ on either side of (\ref{appendix_compare_order_epsilon}) gives
\begin{align}\label{undeformed_hamilton}
    \dot{x}^{i}=\frac{\partial H_{0}}{\partial p^{i}} ( x^{j}, f_{0}^j ) \, ,
\end{align}
which is the Hamilton equation of motion in the undeformed theory, and equating the terms of order $\epsilon$ gives
\begin{align}
    f_{1}^j \frac{\partial^{2}H_{0}}{\partial p^i \partial p^j} + \frac{\partial \fO }{\partial p^{i}} = 0 \, .
\end{align}
This relation describes how $f_1^i$ is implicitly determined in terms of $H_0$ and $\fO$. Since the Legendre transform is an involution, we can express the deformed Lagrangian in terms of the deformed Hamiltonian,
\begin{align}
    L = p^{i} \dot{x}^{i} - H_{0}(x^{i}, p^{i}) -\epsilon \fO ( x^{i}, p^{i} ) \, .
\end{align}
Using the equation for $p^{i}$, we obtain
\begin{align}
    L &= ( f_{0}^i + \epsilon f_{1}^i ) \dot{x}^{i} - H_{0} ( x^{i}, f_{0}^i + \epsilon f_{1}^i ) - \epsilon \fO (x^{i}, f_{0}^i + \epsilon f_{1}^i ) \nonumber \\
    &= f_{0}^i \dot{x}^{i} - H_{0} ( x^{i}, f_{0}^i ) + \epsilon \left( \dot{x}^{i} - \frac{\partial H_0}{\partial p^{i}} ( x^{j}, f_{0}^j ) \right) f_{1}^i - \epsilon \fO ( x^{i}, f_{0}^i ) \, .
\end{align}
By equation (\ref{undeformed_hamilton}), the term proportional to $\epsilon$ must vanish, leading to
\begin{align}
    L = L_{0} - \epsilon \fO \left( x^{i}, f_{0}^i ( \dot{x}^j ) \right) \, .
\end{align}
This is the claim we sought to prove since deforming the Hamiltonian will give rise to the same results as deforming the Lagrangian according to (\ref{first_order_lagrangian_def}) to first order in $\epsilon$.

Since we are only looking at the first order, it does not matter that the relation between $\dot{x}^{i}$ and $p^{i}$ changes. However, as we saw in Theorem \ref{commute_theorem}, one can extend this argument to all orders in the deformation parameter by using the corrected relationship between velocities and conjugate momenta in the deformed theory.

\bibliographystyle{utphys}
\bibliography{master}

\providecommand{\href}[2]{#2}\begingroup\raggedright\begin{thebibliography}{10}

\bibitem{Poland:2022zhe}
D.~Poland and L.~Rastelli, ``{Snowmass Topical Summary: Formal QFT},'' in {\em
  {Snowmass 2021}}.
\newblock 10, 2022.
\newblock \href{http://www.arXiv.org/abs/2210.03128}{{\tt 2210.03128}}.

\bibitem{Zamolodchikov:2004ce}
A.~B. Zamolodchikov, ``{Expectation value of composite field T anti-T in
  two-dimensional quantum field theory},''
\href{http://www.arXiv.org/abs/hep-th/0401146}{{\tt hep-th/0401146}}.

\bibitem{Dubovsky:2012wk}
S.~Dubovsky, R.~Flauger, and V.~Gorbenko, ``{Solving the Simplest Theory of
  Quantum Gravity},'' {\em JHEP} {\bf 09} (2012) 133,
\href{http://www.arXiv.org/abs/1205.6805}{{\tt 1205.6805}}.

\bibitem{Dubovsky:2013ira}
S.~Dubovsky, V.~Gorbenko, and M.~Mirbabayi, ``{Natural Tuning: Towards A Proof
  of Concept},'' {\em JHEP} {\bf 09} (2013) 045,
\href{http://www.arXiv.org/abs/1305.6939}{{\tt 1305.6939}}.

\bibitem{PhysRev.101.453}
L.~Castillejo, R.~H. Dalitz, and F.~J. Dyson, ``Low's Scattering Equation for
  the Charged and Neutral Scalar Theories,'' {\em Phys. Rev.} {\bf 101} (Jan,
  1956) 453--458.

\bibitem{Rosenhaus:2019utc}
V.~Rosenhaus and M.~Smolkin, ``{Integrability and renormalization under $T \bar
  T$},'' {\em Phys. Rev. D} {\bf 102} (2020), no.~6, 065009,
  \href{http://www.arXiv.org/abs/1909.02640}{{\tt 1909.02640}}.

\bibitem{Dey:2021jyl}
A.~Dey and A.~Fortinsky, ``{Perturbative renormalization of the $
  \mathrm{T}\overline{\mathrm{T}} $-deformed free massive Dirac fermion},''
  {\em JHEP} {\bf 12} (2021) 200,
  \href{http://www.arXiv.org/abs/2109.10525}{{\tt 2109.10525}}.

\bibitem{Chakrabarti:2022lnn}
S.~Chakrabarti, A.~Manna, and M.~Raman, ``{Renormalization in $T
  \overline{T}$-deformed nonintegrable theories},'' {\em Phys. Rev. D} {\bf
  105} (2022), no.~10, 106025, \href{http://www.arXiv.org/abs/2204.03385}{{\tt
  2204.03385}}.

\bibitem{Lee:2021iut}
K.-S. Lee, P.~Yi, and J.~Yoon, ``{$ T\overline{T} $-deformed fermionic theories
  revisited},'' {\em JHEP} {\bf 07} (2021) 217,
  \href{http://www.arXiv.org/abs/2104.09529}{{\tt 2104.09529}}.

\bibitem{Lee:2023uxj}
K.-S. Lee and J.~Yoon, ``{$T\overline{T}$ Deformation of $\mathcal{N}=(1,1)$
  Off-Shell Supersymmetry and Partially Broken Supersymmetry},''
  \href{http://www.arXiv.org/abs/2306.08030}{{\tt 2306.08030}}.

\bibitem{Cardy:2018sdv}
J.~Cardy, ``{The $ T\overline{T} $ deformation of quantum field theory as
  random geometry},'' {\em JHEP} {\bf 10} (2018) 186,
\href{http://www.arXiv.org/abs/1801.06895}{{\tt 1801.06895}}.

\bibitem{Datta:2018thy}
S.~Datta and Y.~Jiang, ``{$T\bar{T}$ deformed partition functions},'' {\em
  JHEP} {\bf 08} (2018) 106,
\href{http://www.arXiv.org/abs/1806.07426}{{\tt 1806.07426}}.

\bibitem{Aharony:2018bad}
O.~Aharony, S.~Datta, A.~Giveon, Y.~Jiang, and D.~Kutasov, ``{Modular
  invariance and uniqueness of $T\bar{T}$ deformed CFT},'' {\em JHEP} {\bf 01}
  (2019) 086,
\href{http://www.arXiv.org/abs/1808.02492}{{\tt 1808.02492}}.

\bibitem{Bandos:2020jsw}
I.~Bandos, K.~Lechner, D.~Sorokin, and P.~K. Townsend, ``{A non-linear
  duality-invariant conformal extension of Maxwell's equations},'' {\em Phys.
  Rev. D} {\bf 102} (2020) 121703,
  \href{http://www.arXiv.org/abs/2007.09092}{{\tt 2007.09092}}.

\bibitem{Babaei-Aghbolagh:2022uij}
H.~Babaei-Aghbolagh, K.~B. Velni, D.~M. Yekta, and H.~Mohammadzadeh,
  ``{Emergence of non-linear electrodynamic theories from $T\bar{T}$-like
  deformations},'' \href{http://www.arXiv.org/abs/2202.11156}{{\tt
  2202.11156}}.

\bibitem{Conti:2022egv}
R.~Conti, J.~Romano, and R.~Tateo, ``{Metric approach to a $
  \mathrm{T}\overline{\mathrm{T}} $-like deformation in arbitrary
  dimensions},'' {\em JHEP} {\bf 09} (2022) 085,
  \href{http://www.arXiv.org/abs/2206.03415}{{\tt 2206.03415}}.

\bibitem{Conti:2018jho}
R.~Conti, L.~Iannella, S.~Negro, and R.~Tateo, ``{Generalised Born-Infeld
  models, Lax operators and the $ \mathrm{T}\overline{\mathrm{T}} $
  perturbation},'' {\em JHEP} {\bf 11} (2018) 007,
\href{http://www.arXiv.org/abs/1806.11515}{{\tt 1806.11515}}.

\bibitem{Ferko:2022iru}
C.~Ferko, L.~Smith, and G.~Tartaglino-Mazzucchelli, ``{On Current-Squared Flows
  and ModMax Theories},'' {\em SciPost Phys.} {\bf 13} (2022), no.~2, 012,
  \href{http://www.arXiv.org/abs/2203.01085}{{\tt 2203.01085}}.

\bibitem{Ferko:2023ruw}
C.~Ferko, L.~Smith, and G.~Tartaglino-Mazzucchelli, ``{Stress Tensor Flows,
  Birefringence in Non-Linear Electrodynamics, and Supersymmetry},''
  \href{http://www.arXiv.org/abs/2301.10411}{{\tt 2301.10411}}.

\bibitem{Ferko:2023sps}
C.~Ferko, Y.~Hu, Z.~Huang, K.~Koutrolikos, and G.~Tartaglino-Mazzucchelli,
  ``{$T \overline{T}$-Like Flows and $3d$ Nonlinear Supersymmetry},''
  \href{http://www.arXiv.org/abs/2302.10410}{{\tt 2302.10410}}.

\bibitem{Ferko:2022cix}
C.~Ferko, A.~Sfondrini, L.~Smith, and G.~Tartaglino-Mazzucchelli, ``{Root-$T
  \bar T$ Deformations in Two-Dimensional Quantum Field Theories},'' {\em Phys.
  Rev. Lett.} {\bf 129} (2022), no.~20, 201604,
  \href{http://www.arXiv.org/abs/2206.10515}{{\tt 2206.10515}}.

\bibitem{Rodriguez:2021tcz}
P.~Rodr\'\i{}guez, D.~Tempo, and R.~Troncoso, ``{Mapping relativistic to
  ultra/non-relativistic conformal symmetries in 2D and finite $
  \sqrt{T\overline{T}} $ deformations},'' {\em JHEP} {\bf 11} (2021) 133,
  \href{http://www.arXiv.org/abs/2106.09750}{{\tt 2106.09750}}.

\bibitem{Bagchi:2022nvj}
A.~Bagchi, A.~Banerjee, and H.~Muraki, ``{Boosting to BMS},'' {\em JHEP} {\bf
  09} (2022) 251, \href{http://www.arXiv.org/abs/2205.05094}{{\tt 2205.05094}}.

\bibitem{Tempo:2022ndz}
D.~Tempo and R.~Troncoso, ``{Nonlinear automorphism of the conformal algebra in
  2D and continuous $ \sqrt{T\overline{T}} $ deformations},'' {\em JHEP} {\bf
  12} (2022) 129, \href{http://www.arXiv.org/abs/2210.00059}{{\tt 2210.00059}}.

\bibitem{Hou:2022csf}
J.~Hou, ``{$ T\overline{T} $ flow as characteristic flows},'' {\em JHEP} {\bf
  03} (2023) 243, \href{http://www.arXiv.org/abs/2208.05391}{{\tt 2208.05391}}.

\bibitem{Babaei-Aghbolagh:2022leo}
H.~Babaei-Aghbolagh, K.~Babaei~Velni, D.~Mahdavian~Yekta, and H.~Mohammadzadeh,
  ``{Marginal $T \overline{T}$-like deformation and modified Maxwell theories
  in two dimensions},'' {\em Phys. Rev. D} {\bf 106} (2022), no.~8, 086022,
  \href{http://www.arXiv.org/abs/2206.12677}{{\tt 2206.12677}}.

\bibitem{Borsato:2022tmu}
R.~Borsato, C.~Ferko, and A.~Sfondrini, ``{Classical integrability of root-$T
  \overline{T}$ flows},'' {\em Phys. Rev. D} {\bf 107} (2023), no.~8, 086011,
  \href{http://www.arXiv.org/abs/2209.14274}{{\tt 2209.14274}}.

\bibitem{Ebert:2023tih}
S.~Ebert, C.~Ferko, and Z.~Sun, ``{Root-$T \overline{T}$ Deformed Boundary
  Conditions in Holography},'' \href{http://www.arXiv.org/abs/2304.08723}{{\tt
  2304.08723}}.

\bibitem{Ferko:2023ozb}
C.~Ferko and A.~Gupta, ``{ModMax oscillators and root-TT-like flows in
  supersymmetric quantum mechanics},'' {\em Phys. Rev. D} {\bf 108} (2023),
  no.~4, 046013, \href{http://www.arXiv.org/abs/2306.14575}{{\tt 2306.14575}}.

\bibitem{Garcia:2022wad}
J.~A. Garcia and R.~A. Sanchez-Isidro, ``{$\sqrt{T\overline{T}}$-deformed
  oscillator inspired by ModMax},'' {\em Eur. Phys. J. Plus} {\bf 138} (2023),
  no.~2, 114, \href{http://www.arXiv.org/abs/2209.06296}{{\tt 2209.06296}}.

\bibitem{Bandos:2021rqy}
I.~Bandos, K.~Lechner, D.~Sorokin, and P.~K. Townsend, ``{ModMax meets Susy},''
  {\em JHEP} {\bf 10} (2021) 031,
  \href{http://www.arXiv.org/abs/2106.07547}{{\tt 2106.07547}}.

\bibitem{Kuzenko:2021cvx}
S.~M. Kuzenko, ``{Superconformal duality-invariant models and $ \mathcal{N} $ =
  4 SYM effective action},'' {\em JHEP} {\bf 09} (2021) 180,
  \href{http://www.arXiv.org/abs/2106.07173}{{\tt 2106.07173}}.

\bibitem{Bandos:2020hgy}
I.~Bandos, K.~Lechner, D.~Sorokin, and P.~K. Townsend, ``{On p-form gauge
  theories and their conformal limits},'' {\em JHEP} {\bf 03} (2021) 022,
  \href{http://www.arXiv.org/abs/2012.09286}{{\tt 2012.09286}}.

\bibitem{Kuzenko:2023ysh}
S.~M. Kuzenko and I.~N. McArthur, ``{A supersymmetric nonlinear sigma model
  analogue of the ModMax theory},'' {\em JHEP} {\bf 05} (2023) 127,
  \href{http://www.arXiv.org/abs/2303.15139}{{\tt 2303.15139}}.

\bibitem{Gross:2019ach}
D.~J. Gross, J.~Kruthoff, A.~Rolph, and E.~Shaghoulian, ``{$T\overline{T}$ in
  AdS$_2$ and Quantum Mechanics},'' {\em Phys. Rev. D} {\bf 101} (2020), no.~2,
  026011, \href{http://www.arXiv.org/abs/1907.04873}{{\tt 1907.04873}}.

\bibitem{Gross:2019uxi}
D.~J. Gross, J.~Kruthoff, A.~Rolph, and E.~Shaghoulian, ``{Hamiltonian
  deformations in quantum mechanics, $T\bar T$, and the SYK model},'' {\em
  Phys. Rev. D} {\bf 102} (2020), no.~4, 046019,
  \href{http://www.arXiv.org/abs/1912.06132}{{\tt 1912.06132}}.

\bibitem{Lechner:2022qhb}
K.~Lechner, P.~Marchetti, A.~Sainaghi, and D.~P. Sorokin, ``{Maximally
  symmetric nonlinear extension of electrodynamics and charged particles},''
  {\em Phys. Rev. D} {\bf 106} (2022), no.~1, 016009,
  \href{http://www.arXiv.org/abs/2206.04657}{{\tt 2206.04657}}.

\bibitem{Ferko:2021loo}
C.~Ferko, {\em {Supersymmetry and Irrelevant Deformations}}.
\newblock PhD thesis, Chicago U., 2021.
\newblock \href{http://www.arXiv.org/abs/2112.14647}{{\tt 2112.14647}}.

\bibitem{Ebert:2022xfh}
S.~Ebert, C.~Ferko, H.-Y. Sun, and Z.~Sun, ``{$T \overline{T}$ Deformations of
  Supersymmetric Quantum Mechanics},''
  \href{http://www.arXiv.org/abs/2204.05897}{{\tt 2204.05897}}.

\bibitem{Kruthoff:2020hsi}
J.~Kruthoff and O.~Parrikar, ``{On the flow of states under $T\overline{T}$},''
  \href{http://www.arXiv.org/abs/2006.03054}{{\tt 2006.03054}}.

\bibitem{Brennan:2019azg}
T.~D. Brennan, C.~Ferko, and S.~Sethi, ``{A Non-Abelian Analogue of DBI from $T
  \overline{T}$},'' {\em SciPost Phys.} {\bf 8} (2020), no.~4, 052,
  \href{http://www.arXiv.org/abs/1912.12389}{{\tt 1912.12389}}.

\bibitem{Matsoukas-Roubeas:2022odk}
A.~S. Matsoukas-Roubeas, F.~Roccati, J.~Cornelius, Z.~Xu, A.~Chenu, and A.~del
  Campo, ``{Non-Hermitian Hamiltonian deformations in quantum mechanics},''
  {\em JHEP} {\bf 01} (2023) 060,
  \href{http://www.arXiv.org/abs/2211.05437}{{\tt 2211.05437}}.

\bibitem{Chakraborty:2020xwo}
S.~Chakraborty and A.~Mishra, ``{$ T\overline{T} $ and $ J\overline{T} $
  deformations in quantum mechanics},'' {\em JHEP} {\bf 11} (2020) 099,
  \href{http://www.arXiv.org/abs/2008.01333}{{\tt 2008.01333}}.

\bibitem{CoppaThesis}
G.~Coppa, {\em {Elettrodinamica non lineare di Born e Infeld}}.
\newblock {Tesi di Laurea Magistrale in Fisica, Università degli Studi di
  Torino}, {2019}.

\bibitem{Giordano:2023byh}
F.~Giordano, S.~Negro, and R.~Tateo, ``{The Generalised Born Oscillator and the
  Berry-Keating Hamiltonian},'' \href{http://www.arXiv.org/abs/2307.15025}{{\tt
  2307.15025}}.

\bibitem{Iliesiu:2020zld}
L.~V. Iliesiu, J.~Kruthoff, G.~J. Turiaci, and H.~Verlinde, ``{JT gravity at
  finite cutoff},'' {\em SciPost Phys.} {\bf 9} (2020) 023,
  \href{http://www.arXiv.org/abs/2004.07242}{{\tt 2004.07242}}.

\bibitem{Ali:2004ft}
S.~T. Ali and M.~Englis, ``{Quantization methods: A Guide for physicists and
  analysts},'' {\em Rev. Math. Phys.} {\bf 17} (2005) 391--490,
  \href{http://www.arXiv.org/abs/math-ph/0405065}{{\tt math-ph/0405065}}.

\bibitem{Ramond:1981pw}
P.~Ramond, {\em {Field Theory: A Modern Primer}}.
\newblock {Second}~ed., {2001}.

\bibitem{ZinnJustin2005PathII}
J.~Zinn-Justin, {\em Path integrals in quantum mechanics}.
\newblock 2005.

\bibitem{Mosel:2004mk}
U.~Mosel, {\em {Path integrals in field theory: An introduction}}.
\newblock 2004.

\bibitem{Ferko:2022dpg}
C.~Ferko and S.~Sethi, ``{Sequential flows by irrelevant operators},'' {\em
  SciPost Phys.} {\bf 14} (2023), no.~5, 098,
  \href{http://www.arXiv.org/abs/2206.04787}{{\tt 2206.04787}}.

\bibitem{Ferko:2023wyi}
C.~Ferko, S.~M. Kuzenko, L.~Smith, and G.~Tartaglino-Mazzucchelli,
  ``{Duality-Invariant Non-linear Electrodynamics and Stress Tensor Flows},''
  \href{http://www.arXiv.org/abs/2309.04253}{{\tt 2309.04253}}.

\bibitem{Russo:2022qvz}
J.~G. Russo and P.~K. Townsend, ``{Nonlinear electrodynamics without
  birefringence},'' {\em JHEP} {\bf 01} (2023) 039,
  \href{http://www.arXiv.org/abs/2211.10689}{{\tt 2211.10689}}.

\bibitem{olver10}
F.~W.~J. Olver, W.~L. Daniel, R.~F. Boisvert, and C.~W. Clark, {\em The {NIST}
  Handbook of Mathematical Functions}.
\newblock Cambridge Univ. Press, 2010.

\bibitem{10.1119/1.17809}
H.~Haugerud and F.~Ravndal, ``{The two‐dimensional harmonic oscillator at
  finite temperature and nonzero chemical potential},'' {\em American Journal
  of Physics} {\bf 63} (09, 1995) 839--844,
  \href{http://www.arXiv.org/abs/https://pubs.aip.org/aapt/ajp/article-pdf/63/9/839/12163256/839\_1\_online.pdf}{{\tt
  https://pubs.aip.org/aapt/ajp/article-pdf/63/9/839/12163256/839\_1\_online.pdf}}.

\bibitem{Grassi:2018bci}
A.~Grassi and M.~Mari\~no, ``{A Solvable Deformation of Quantum Mechanics},''
  {\em SIGMA} {\bf 15} (2019) 025,
  \href{http://www.arXiv.org/abs/1806.01407}{{\tt 1806.01407}}.

\bibitem{Datta:2021ftn}
S.~Datta, S.~Duary, P.~Kraus, P.~Maity, and A.~Maloney, ``{Adding flavor to the
  Narain ensemble},'' {\em JHEP} {\bf 05} (2022) 090,
  \href{http://www.arXiv.org/abs/2102.12509}{{\tt 2102.12509}}.

\end{thebibliography}\endgroup

\end{document}